\documentclass[a4paper,oneside,10pt]{article}

\usepackage[paperwidth=171mm,paperheight=246mm,top=25mm,bottom=25mm, left=14mm, right=14mm]{geometry} 

\usepackage[english]{babel}

\usepackage{graphicx}
\usepackage[font=small,labelfont=bf]{caption}
\usepackage{subcaption}
\usepackage{color}
\usepackage{mdframed}
\usepackage{framed} 
\usepackage{cancel}
\usepackage[normalem]{ulem}

\usepackage{multicol}

\usepackage{tikz}
\usetikzlibrary{tikzmark}

\usepackage{bookmark}
\usepackage{cancel}
\usepackage{multirow}
\usepackage{hhline}
\usepackage{diagbox}
\usepackage{makecell}
\usepackage{slashbox,booktabs}

\usepackage{url}

\usepackage{enumerate}
\usepackage{here}

\usepackage[normalem]{ulem}
\usepackage{ragged2e}
\usepackage{csquotes}
\usepackage{tasks}
\usepackage{paralist}
\usepackage[inline]{enumitem}
\usepackage{tikz-cd}
\usepackage{stmaryrd}

\usepackage{afterpage}
\usepackage{lineno,hyperref}
\modulolinenumbers[5]

\usepackage{amsmath,amssymb,amsthm,mathtools,mathrsfs,epsfig,amsfonts,wasysym}
\usepackage{MnSymbol}

\usepackage[normalem]{ulem}
\usepackage{ragged2e}
\usepackage{csquotes}
\usepackage{tasks}
\usepackage{paralist}
\usepackage[inline]{enumitem}
\usepackage{tikz-cd}
\usepackage{stmaryrd}

\usepackage{caption}

\usepackage[titletoc,title]{appendix}

\newcommand{\de}{\mathrm{d}}

\DeclareMathOperator*{\fiint}{\ensuremath{\iint\text{\kern-1.7em{\raisebox{3.5pt}{\rotatebox{-82}{$\big|$}}}}}}

\def\gr{\operatornamewithlimits{\mathrm{gr}}}

\usepackage{url,hyperref}
\hypersetup{colorlinks=true,linkcolor=blue,filecolor=magenta,urlcolor=cyan,citecolor=Peru}
\definecolor{Peru}{rgb}{0.72,0.48,0.24}

\usepackage{nicematrix}
\usepackage{tikz}
\usetikzlibrary{fit}

\usepackage[all]{xy}

\usepackage{booktabs}
\usepackage{tabularray}
\UseTblrLibrary{diagbox}
\usepackage{colortbl}

\usepackage{authblk}
\usepackage{todonotes}

\definecolor{wildstrawberry}{rgb}{1.0, 0.26, 0.64}
\newcommand{\vb}[1]{\textcolor{wildstrawberry}{#1}}

\definecolor{ao(english)}{rgb}{0.0, 0.5, 0.0}

\usepackage{physics}
\usepackage{xparse}
\DeclarePairedDelimiterX{\set}[1]{\{}{\}}{\setargs{#1}}
\NewDocumentCommand{\setargs}{>{\SplitArgument{1}{|}}m}
{\setargsaux#1}
\NewDocumentCommand{\setargsaux}{mm}
{\IfNoValueTF{#2}{#1} {#1\,\delimsize|\,\mathopen{}#2}}
\usepackage{nicematrix}

\usepackage{orcidlink}

\newcommand{\reff}[2]{\hyperref[{#2}]{{#1} \ref*{#2}}}

\begin{document}

\title{From primary HPV infection to carcinoma \textit{in situ}: a mathematical approach of cervical intraepithelial neoplasia} 

\author[1]{Vasiliki Bitsouni\,\orcidlink{0000-0002-0684-0583}\thanks{\texttt{vbitsouni@math.upatras.gr}}}

\author[2,3]{Nikolaos Gialelis\,\orcidlink{0000-0002-6465-7242}\thanks{\texttt{ngialelis@math.uoa.gr}}}

\author[2]{Ioannis G. Stratis\,\orcidlink{0000-0002-0179-0820}\thanks{\texttt{istratis@math.uoa.gr}}}

\author[1]{Vasilis Tsilidis\,\orcidlink{0000-0001-5868-4984}\thanks{\texttt{vtsilidis@upatras.gr}}}

\affil[1]{Department of Mathematics, University of Patras, GR-26504 Rio Patras, Greece}

\affil[2]{Department of Mathematics, National and Kapodistrian University of Athens, GR-15784 Athens, Greece}

\affil[3]{School of Medicine, National and Kapodistrian University of Athens,\newline GR-11527 Athens, Greece}

\date{}

\maketitle

\begin{abstract}
\noindent
Cervical intraepithelial neoplasia (CIN) is the development of abnormal cells on the surface of the cervix, caused by a human papillomavirus (HPV) infection. Although in most of the cases it is resolved by the immune system, a small percentage of people might develop a more serious CIN which, if left untreated, can develop into cervical cancer. Cervical cancer is the fourth most common cancer in women globally, for which the World Health Organization (WHO) recently adopted the Global Strategy for cervical cancer elimination by 2030. With this research topic being more imperative than ever, in this paper, we develop a nonlinear mathematical model describing the CIN progression. The model consists of partial differential equations describing the dynamics of epithelial, dysplastic and immune cells, as well as the dynamics of viral particles. We use our model to explore numerically three  important factors of dysplasia progression, namely the geometry of the cervix, the strength of the immune response and the frequency of viral exposure.
\end{abstract}


\section{Introduction}
\label{intro}

Healthy cells multiply and die in a orderly way, so that each new replaces one lost.  However, if a cell is damaged, e.g. due to a virus infection that the immune system failed to defeat, the new damaged cells will either die or start to proliferate in an uncontrolled manner, creating a signalling of oncogenes that act by mimicking growth signalling \cite{hanahan2000hallmarks}. The alterations in signal pathways, which can impact the cell's normal biological behavior, include changes such as excessive proliferation, resistance to apoptosis, and evasion of immune response, leading to the development of an abnormal mass of tissue that differs in clinically important phenotypic features; the tumour \cite{marusyk2012intra}. The most common cancers develop from the skin, breast, endometrium, prostate, colon, lung, pancreas, bladder, liver, and cervix \cite{mendez2022revisiting}. Organs and blood vessels throughout the body are covered by a protective layer of compactly packed cells with a little intercellular matrix, the epithelium or epithelial tissue.  These cells, known as epithelial cells, may undergo the aforementioned abnormality which may result in cancer. 

A key role in cancer growth, as well as diagnosis and treatment \cite{ogino2011cancer}, is played by the immune system, a complex network of organs, cells and proteins that defends the body against infection that bacteria, viruses, fungi or parasites can cause while protecting the body’s own cells. It is a collection of reactions and responses that the body makes to damaged cells or infections. So, it is sometimes called the immune response. The immune system is important to people with cancer because: $(\mathrm{i})$ cancer can weaken the immune system; $(\mathrm{ii})$ cancer treatments might weaken the immune system; and $(\mathrm{iii})$ the immune system may help to fight cancer.

Cervical cancer is the fourth most common cancer in women globally with an estimated 604 000 new cases and 342 000 deaths in 2020 \cite{Who2023cc}, representing nearly 8\% of all female cancer deaths every year \cite{sung2021global}. About 90\% of these deaths caused by cervical cancer occurred in low- and middle-income countries with sub-Saharan Africa (SSA), Central America and South-East Asia having the highest rates of cervical cancer incidence and mortality \cite{Who2023cc}. 

World Health Organization (WHO) has classified the premalignant lesions mild, moderate, or severe dysplasia or carcinoma \textit{in situ} (CIS) \cite{demay1999practical}. Carcinoma \textit{in situ}, a.k.a. stage 0 cancer or `\textit{in situ} neoplasm', is the stage at which a group of abnormal cells in an area of the body appears. The cells may develop into cancer at some time in the future. The changes in the cells are called dysplasia and at this stage the number of abnormal cells is too small to form a tumour. These cell changes, a.k.a. `precancerous changes' or `non-invasive cancer', may not develop into cancer, and usually the carcinoma \textit{in situ} is too small to show up on a scan. 

In order to emphasize the spectrum of abnormality in these lesions, and to help standardize treatment, the term cervical intraepithelial neoplasia (CIN) was developed \cite{santesso2016world}. For premalignant dysplastic changes, cervical intraepithelial neoplasia grading (CIN1--3) is used, classifying mild dysplasia as CIN1, moderate dysplasia as CIN2, and severe dysplasia and CIS as CIN3 \cite{salcedo2021intraepithelial}. Histologically the epithelial tumours of the uterine cervix can be classified into squamous cell carcinoma (SCC), glandular tumours and precursors, mesenchymal tumours and tumour-like conditions, mixed epithelial and mesenchymal tumours, melanocytic tumours, miscellaneous tumours, lymphoid and haematopoietic tumours, and secondary tumours \cite{Whohc}. From the aforementioned types of cervical cancers, the most commonly reported are SCC (75\%) and adenocarcinoma, a type of the glandular tumour and precursors (25\%). 

It has been reported that approximately 95\% of cervical cancer cases are caused by persistent genital high-risk human papillomavirus (HPV) infection \cite{schiffman2007human,guan2012human,intgen2017,kusakabe2023carcinogenesis}. HPV is a common sexually transmitted infection (STI), which can affect the skin, genital area and throat \cite{bzhalava2013systematic,tommasino2014human}. All sexually active people have been susceptible to the infections once in a lifetime, regardless of gender, genetic background, and geographical location. 90\% of the cases are being resolved with no symptoms within two years \cite{Who2023cc}, whereas in some cases, the infection persists and results in either warts or precancerous lesions \cite{ljubojevic2014hpv}. There are nearly 200 types of HPV \cite{nci2023hpv}. Types 16 (reported as the most carcinogenic and associated with more than 60\% of  cervical squamous cancers and adenocarcinomas \cite{perkins2023cervical}) and 18 (more commonly associated with  cervical squamous cancers and adenocarcinomas, and together with type 45 cause approximately 20\% of cancers \cite{perkins2023cervical}) are referred to as high-risk and are being detected in more than 90\% of cervical cancer lesions \cite{walboomers1999human}.

The life cycle of HPV is strongly linked to the differentiation state of the host epithelial cell and is governed by the action of both viral and cellular proteins \cite{moody2010human,doorbar2015human}. First, the virus accesses  the basal epithelial cells in the cervix of a woman through microlesions (see \reff{Figure}{fig:CIN}). By entering these basal cells of the squamous epithelium, the virus establishes the viral cycle. Then, through bindings the viral material manages to enter into the nucleus, where the virus deploys the host cell replication machinery and starts viral genomic replication at about 50--200 copies per cell \cite{stubenrauch1999human,keiffer2021recent}.  The infected cells either remain persistently infected until they become cleared by the immune response or progress to cancers. In particular, the infected cells express viral proteins interacting with the normal cell cycle, promoting proliferation and deactivating certain tumour suppressor proteins \cite{asih2016dynamics}. It follows a viral genome replication and cell division that would lead in migration of the daughter cells from the basal layer to the other layers, as well as an uncontrolled proliferation of the HPV-infected cells \cite{lee2007differentiation}. 

Since the 1920s when the Papanicolaou test (also referred to as Pap test, Pap smear, cervical smear, cervical screening) was introduced by the Greek physician George N. Papanicolaou as a cervical screening method used to detect potentially precancerous and cancerous processes in the cervix, there has been ongoing research not only for the detection, but most importantly for the prevention of HPV infection and as a consequence of cervical cancer. Tests used in cancer screening programs can pick up carcinomas \textit{in situ} in the cervix. In fact, a very recent cervical screening self-test is being introduced in more and more countries within the  Global Strategy for cervical cancer elimination by 2030 \cite{Who2030}, which proposes global vaccination, screening and treatment. As for the prevention, the first HPV vaccine became available in 2006 and since then 125 countries include HPV vaccines in their routine vaccinations for girls, and 47 countries also for boys \cite{mondiale2022human}. Currently there are six licensed HPV vaccines, which are highly efficacious in preventing infection with virus types 16 and 18 \cite{Who2022vaccine}. Prophylactic vaccination against HPV, screening and treatment of precancer lesions are effective ways to prevent cervical cancer and are very cost-effective. If cervical cancer is diagnosed at an early stage and treated promptly it can be cured. However, the first step to be taken is to understand the mechanism between the HPV infection and cervical cancer progression depending on the $(\mathrm{i})$ viral load, $(\mathrm{ii})$ geometry of the under-examination domain and $(\mathrm{iii})$ immune response.
\begin{figure}[ht]
\centering
\includegraphics[width=1\textwidth]{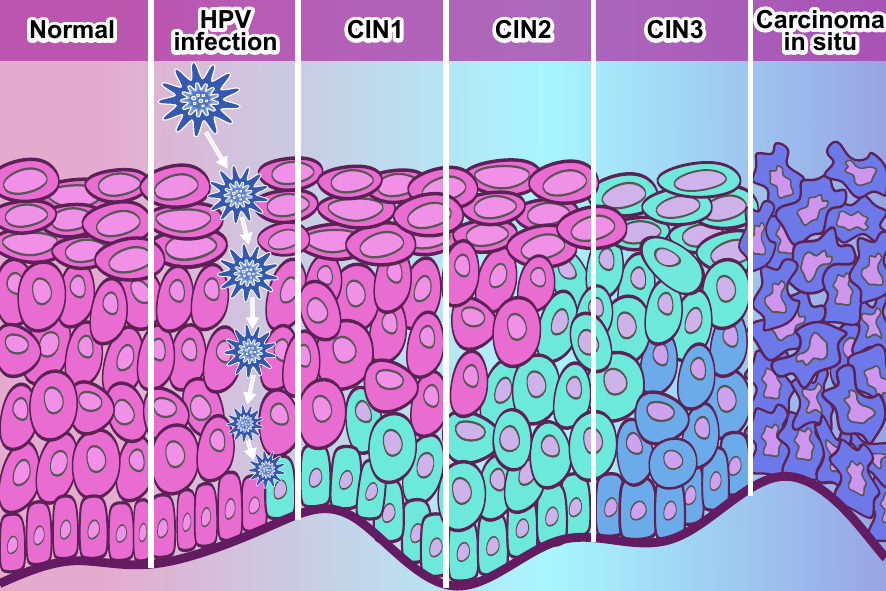} 
\caption{Cervical precancerous lesions classification based on the CIN (cervical intraepithelial neoplasia) staging. These lesions are initiated with HPV infection. HPV viral particles traverse the epithelium, infecting the cells of the basal layer.}
\label{fig:CIN}
\end{figure}

To the authors' best knowledge, the literature about precancerous lesions that lead to carcinoma \textit{in situ} is poor (see, e.g. \cite{solis2017numerical,solis2023nonlinear}). The early mathematical models on cervical cancer were focused on epidemiology, describing the transmission dynamics between individuals and the impact of the HPV vaccine (see e.g. \cite{barnabas2006epidemiology,elbasha2008global,brown2010hpv,brown2011role,lee2012mathematical,sado2019mathematical,gurmu2020mathematical,rajan2023mathematical} and many references therein). There are only a few later examples of mathematical models for the dynamics of HPV-infected cells at the molecular or tissue levels, such as the models in \cite{asih2016dynamics,chakraborty2019role}, which were focused on the progression of cervical cells from normal cells into precancerous and cancerous classes. 

In the present work, a continuum mechanical model is introduced in order to describe the spatiotemporal dynamics of the cervical epithelial cells, HPV viral particles, dysplastic and immune cells, and the cervical intraepithelial neoplasia progression is studied as a result of their interaction. The focus of this work is given to certain, key---as it is established here---factors of the dysplasia progression, such as the geometry of the basement membrane, the strength of the immune response and the frequency of viral exposure. To this end, various scenarios will be studied and 3D numerical simulations will be run in order to reproduce the patterns widely accepted by the medical community concerning the staging of the lesions.

The layout of this paper is as follows. In \hyperref[model-form]{\S \ref*{model-form}} we formulate a novel model of nonlinear partial differential equations for the dynamics of the epithelial, dysplastic and immune cells, as well as the viral particles. In \hyperref[scenarios]{\S \ref*{scenarios}} we numerically investigate certain key factors of dysplasia progression. We conclude in \hyperref[conclusion]{\S \ref*{conclusion}} with a summary and discussion of the results.
\section{Model formulation} 
\label{model-form}

In this section, we develop the spatial and dynamical parts of our model. 

\subsection{Histology and spatial structure}
\label{sec:histology}

The vaginal epithelium is a type of stratified squamous epithelium that lines the vagina. It plays a significant role in protecting the body from pathogens. The endocervix epithelium is a type of columnar epithelium that lines the endocervix. The transitional area between the two aforementioned epithelia is called squamocolumnar junction. This area is susceptible to HPV infection and is the region where physiological transformation to squamous metaplasia occurs \cite{prendiville2017colposcopy}.

Epithelial tissue proliferates or regrows through a process called epithelialization. During epithelialization, epithelial cells, especially keratinocytes, proliferate in the basal layer, differentiate as they rise through the spinous and granular layer, and then lose their nucleus and flatten to become the outer layer of skin known as the \textit{stratum corneum} \cite{pastar2014epithelialization}. 

Hence, we can differentiate the vaginal epithelium into two layers, based on whether the epithelial cells of that layer are proliferating or not; the proliferative layer and the non-proliferative layer.


\subsubsection{Proliferative layer}
\label{l1}

The proliferative layer consists of the basal cells which are supported by the basement membrane. 

\paragraph{Basement membrane}
\label{basement membrane}

The basement membrane is a specialized extracellular matrix structure that separates the epithelium from the underlying connective tissue. It's a thin, pliable sheet-like structure, with its primary function being to anchor the epithelium to the underlying connective tissue. It also provides physical and structural support to the epithelium \cite{kierszenbaum2015histology,pozzi2017nature}.

The shape of the basement membrane is wave-like with neither a specific spatial frequency or amplitude throughout the tissue, as can be observed from various histological images (see \cite{kurman2013blaustein,mescher2018junqueira} and references therein). Hence, in order to investigate how its shape affects the formation of dysplastic cells, we have to be able to generate a wide range of different basement membrane shapes. We will model the basement membrane as a surface, due to it being very thin.

There are many ways to generate a random surface \cite{barnsley1988science}. Due to the wave-like shape of the basement membrane, we will use the method presented in \cite{sjodin2017generate}, as it utilizes the Fourier series---a natural way to describe wave phenomena. Thus, according to \cite{sjodin2017generate}, a random 3D surface can be expressed as:
\begin{equation*} 
    f\qty(x,y) = \sum_{m=-M}^M \sum_{n=-N}^N a\qty(m,n) \cos(\qty(2\pi x,2\pi y) \vdot \vb*{\nu} - \phi\qty(m,n)) \;,
\end{equation*}
where for the $\qty(m,n)$-th term of the Fourier series: $a\qty(m,n)$ is its \textit{amplitude}, $\vb*{\nu}$ is its \textit{spatial frequency}, and $\phi\qty(m,n)$ is its \textit{phase angle}.

Regarding $\vb*{\nu}$, we allow for a discrete set of spatial frequencies:
\begin{equation*}
    \vb*{\nu} = \qty(m,n) \;, \; \text{for}\; \qty(m,n) \in \set*{ \qty(m,n) \in \mathbb{Z}^2 | \qty(m,n) \in \qty[-M,M] \cross \qty[-N,N] } \;.
\end{equation*}

Regarding $\phi$, we assume that each phase angle follows a uniform distribution on the 2D set $\qty[-\frac{\pi}{2},\frac{\pi}{2}]^2$:
\begin{equation*}
    \phi \sim \mathcal{U}\qty( \qty[-\frac{\pi}{2},\frac{\pi}{2}]^2) \;.
\end{equation*}

Regarding $a$, because slower oscillations are more likely to have a larger amplitude than faster ones, we have to reduce the amplitude of the terms of the Fourier series which have high frequency: 
\begin{equation*}
    a\qty(m,n) = a_c(m,n) \ell\qty(m,n) \;,
\end{equation*}
with
\begin{equation*}
    a_c(m,n) = \begin{cases}
        0\;, & \text{for $(m,n) = \vb{0}_2$} \\
        \norm{\vb*{\nu}}^{-\beta} = \qty(m^2 + n^2)^{-\frac{\beta}{2}}\;, & \text{otherwise}\;,
    \end{cases}
\end{equation*}
where $\beta \in \mathbb{R}$ is called the \textit{spectral exponent} and indicates how quickly higher frequencies are attenuated, whereas $\ell$ follows the standard 2D normal distribution:
\begin{equation*}
    \ell \sim \mathcal{N}_2\qty( \vb{0}_2, \vb*{I}_2) \;,
\end{equation*}
where, $\vb{0}_2$ is the zero vector of $\mathbb{R}^2$ and $\vb*{I}_2$ is the identity matrix of $\mathbb{R}^{2\times2}$.
If the the spectral exponent is large (resp. small), higher frequencies will (resp. not) be attenuated, thus making the surface smoother (resp. rougher).

Therefore, the basement membrane is approximated by
\begin{equation*} \label{eq:basal_layer}
        f\qty(x,y) = \sum_{m=-M}^M \sum_{n=-N}^N a_c(m,n) \ell\qty(m,n) \cos( 2\pi \qty(x m + y n)  - \phi\qty(m,n)) \;.
\end{equation*}

\reff{Figure}{fig:random_basement_membrane} depicts different basement membranes for different values of $\beta$ and different samples of $\ell$ and $\phi$.

\begin{figure}[!ht]
    \begin{subfigure}[t]{0.48\textwidth}
        \centering
        \includegraphics[height = 16cm]{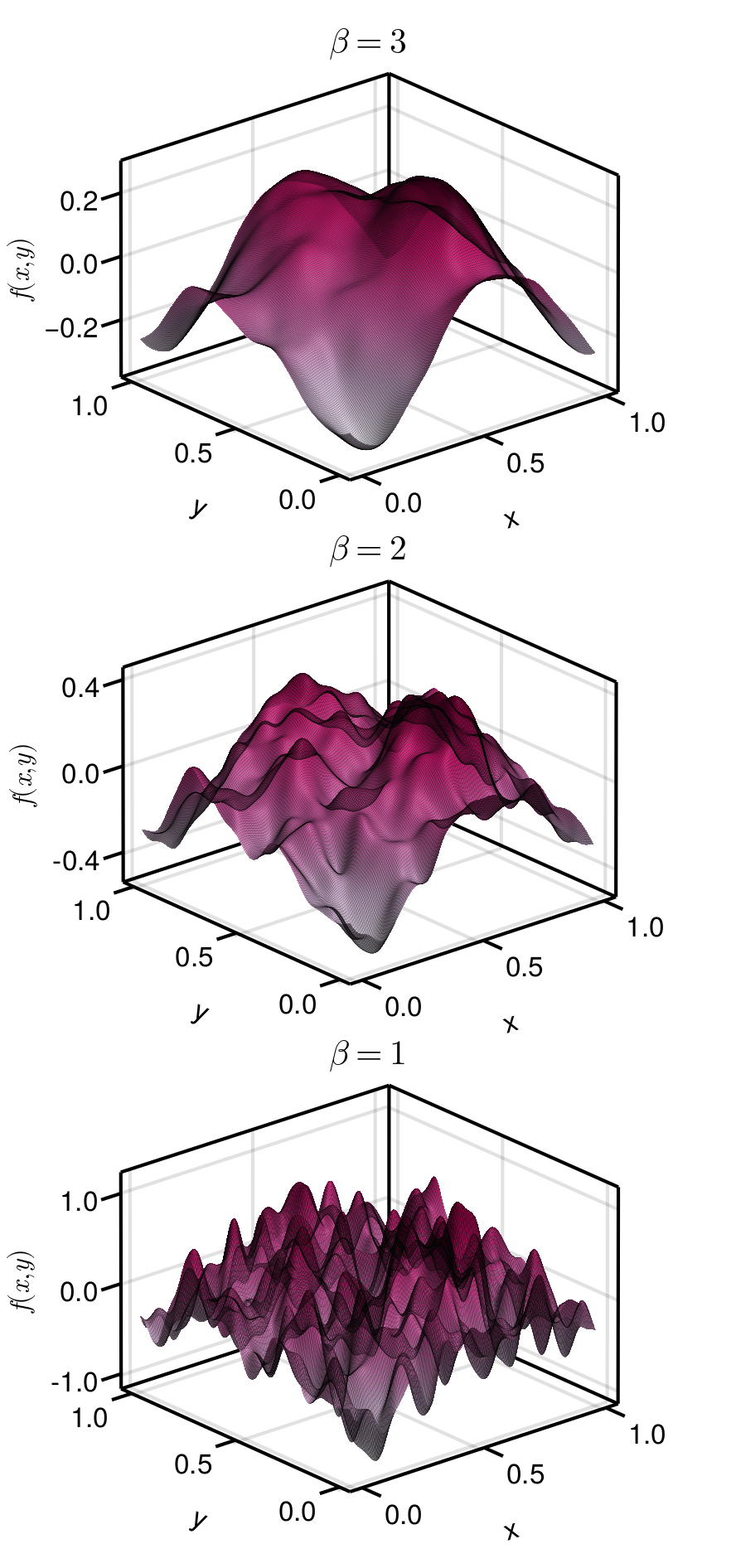}
        \caption{As the  the value of $\beta$ decreases, the  surface becomes rougher.}
        \label{fig:random_basement_membrane_A}
    \end{subfigure}
    \hspace*{\fill}%
    \begin{subfigure}[t]{0.48\textwidth}
        \centering
        \includegraphics[height = 16cm]{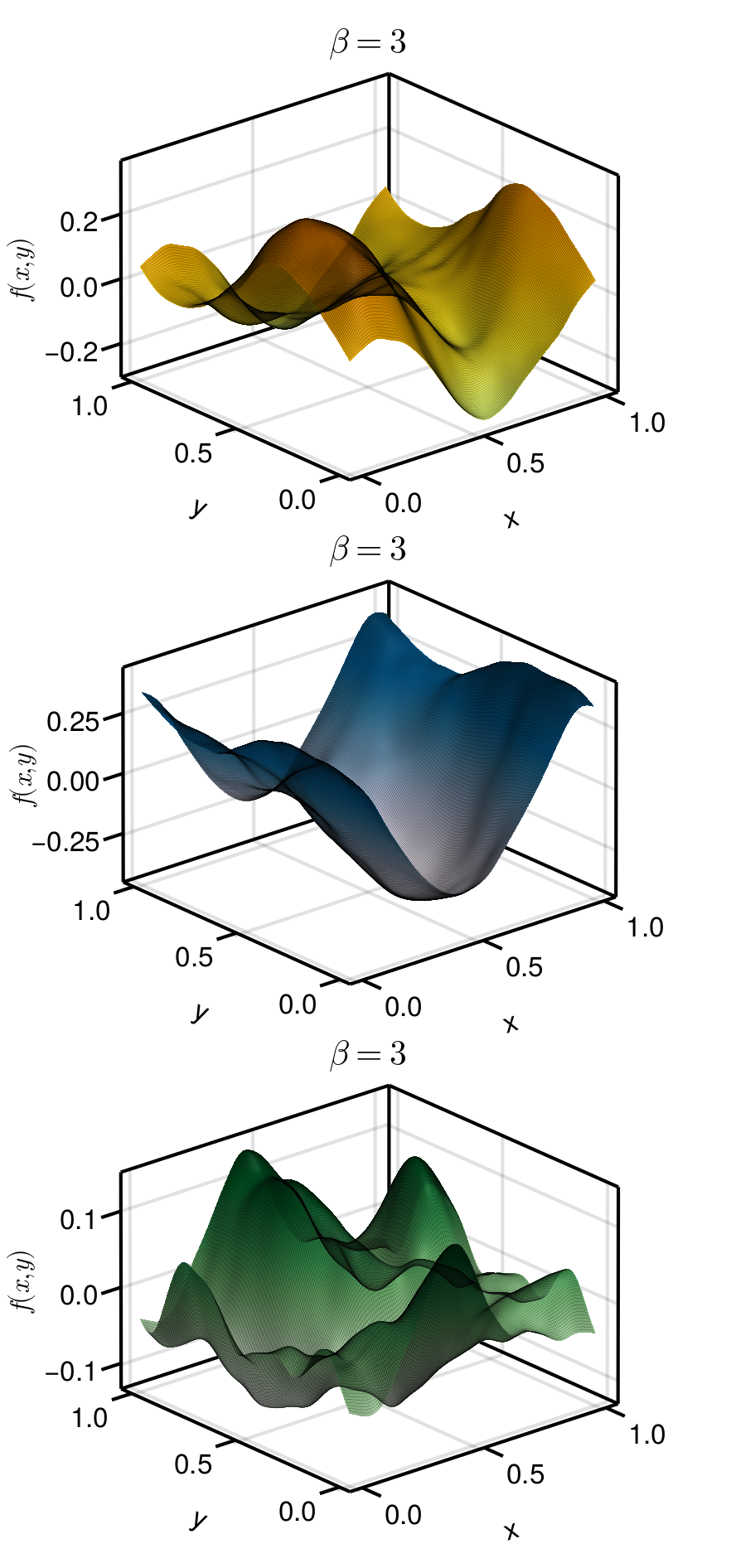}
        \caption{Three different random surfaces with the same value of $\beta$.}
        \label{fig:random_basement_membrane_Β}
    \end{subfigure}
    \caption{The graph of $f$ for random samples of $\ell$ and $\phi$, with (a) different values of $\beta$, and (b) the same value of $\beta$. The different colors represent different pairs of $\ell$ and $\phi$. Additionally, we have that $N=M=10$.}
    \label{fig:random_basement_membrane}
\end{figure} 

\paragraph{Basal layer}
\label{basal layer}

The basal layer, also known as \textit{stratum basale}, is the innermost layer of the epidermis. This layer plays a crucial role in cell replacement and differentiation. It is the layer from which new cells are derived and it is constantly dividing to replace old, damaged cells in the epidermis \cite{kierszenbaum2015histology}. 

We have already seen how to approach the construction of the basement membrane, $f$, as a random 3D surface, but how about the basal layer? This time, things should be specific and not random. In fact, we can rely on certain ubiquitous laws in order to determine the basal layer's geometry. 

Considering that the basal layer lies on a flat basement membrane, $f_0$, then, under the hypothesis that the density is everywhere constant, we expect that the width of the layer, $h$, is also everywhere constant, say $h_0$, and the upper boundary of the layer, $g$, is also a flat graph of constant amplitude, $g_0=f_0+h_0$. In case where $f$ is not constant, a transition between those two cases, which is depicted in \reff{Figure}{fig:toplayer}, should obey the law of conservation of mass, $m$. Hence $$\int\limits_{\substack{\text{arbitrary part of}\\\text{basal layer}\\\text{in }1\text{st case }}}{\de m_0}=\int\limits_{\substack{\text{same part of}\\\text{basal layer}\\\text{in }2\text{nd case }}}{\de m}\;,$$ thus $$\varpi_0\int\limits_{\substack{\text{arbitrary part of}\\\text{basal layer}\\\text{in }1\text{st case }}}{\de V_0}=\varpi_0\int\limits_{\substack{\text{same part of}\\\text{basal layer}\\\text{in }2\text{nd case }}}{\de V}\;,$$ where $\varpi$ and $V$ stand for the density and the volume, respectively, of the layer, or else $$h_0\int\limits_{\substack{\text{arbitrary part of}\\\text{basement membrane}\\\text{in }1\text{st case }}}{\de\sigma_0}=\int\limits_{\substack{\text{same part of}\\\text{basement membrane}\\\text{in }2\text{nd case }}}{h\de\sigma}\;,$$ where $\sigma$ stands for the surface area of the membrane $f$. Taking into consideration the geometry of the problem, we deduce that $$h_0\iint\limits_{D}{\de x\de y}=\iint\limits_{D}{h\sqrt{1+{\left(\frac{\partial f}{\partial x}\right)}^2+{\left(\frac{\partial f}{\partial y}\right)}^2}\de x\de y}\;,$$ or else
$$h_0=\fiint\limits_{D}{h\sqrt{1+{\left(\frac{\partial f}{\partial x}\right)}^2+{\left(\frac{\partial f}{\partial y}\right)}^2}\de x\de y}\;,$$ where $D\subseteq\mathbb{R}^2$ is an arbitrary nontrivial open subset of the domain of $f$.


\begin{figure}[H]
\centering
\includegraphics[width=1\textwidth]{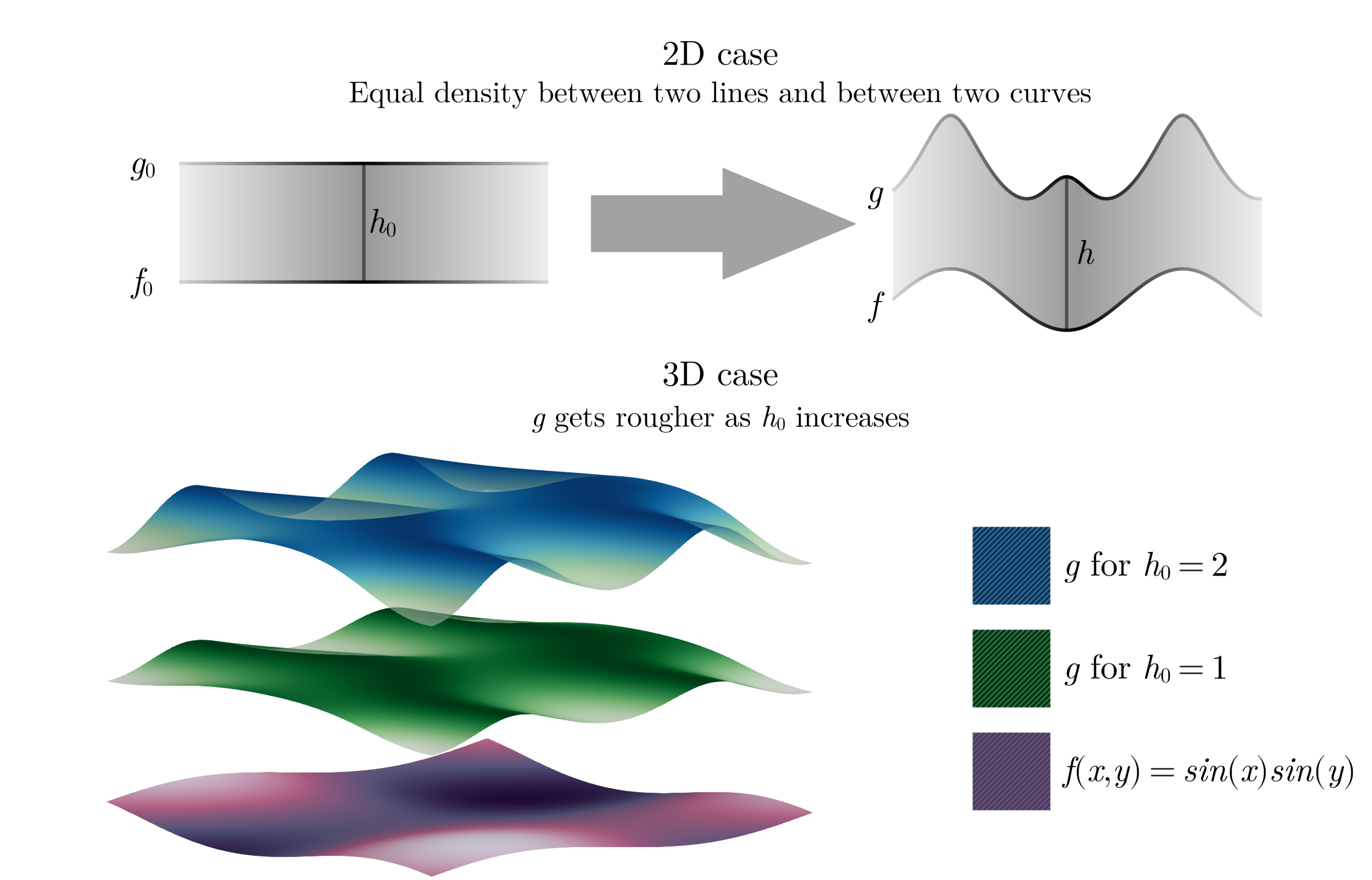} 
\caption{Vertical illustration of the geometry of the basal layer.}
\label{fig:toplayer}
\end{figure}

Now letting $\left|D\right|\to 0$, we get from the Lebesgue-Besicovitch differentiation theorem that
$$h=\frac{h_0}{\sqrt{1+{\left(\frac{\partial f}{\partial x}\right)}^2+{\left(\frac{\partial f}{\partial y}\right)}^2}}\;,$$ therefore $$g=f+\frac{h_0}{\sqrt{1+{\left(\frac{\partial f}{\partial x}\right)}^2+{\left(\frac{\partial f}{\partial y}\right)}^2}}\;.$$

Henceforward, we write $\Omega_3$ for the basal layer, which lies between the surfaces $f$ and $g$.

\subsubsection{Non-proliferative layer}
\label{l2}

Non-proliferative layer is composed of stratified squamous cells and consists of four sublayers: the \textit{stratum granulosum}, \textit{stratum spinosum}, \textit{stratum lucidum}, and \textit{stratum corneum}. Each layer plays a specific role in the epithelium's function and is the result of the squamous cell differentiation program. For example, after a slow coordinated process in space and time, the \textit{stratum corneum} is formulated by a layer of dead cells (corneocytes) to create a physical barrier for the skin.

Henceforth, we write $\Omega_2$ for the non-proliferative layer, which lies on top of $\Omega_3$. 

\subsubsection{Unified domain}
\label{l3}

For modelling needs, we consider $\Omega_1$ to be an additional layer above $\Omega_2$, in order to incorporate the external environment of the epidermis. Moreover, $\Omega$ stands for the union of the three layers, $\Omega_1$, $\Omega_2$ and $\Omega_3$, along with their boundaries, and is presented in \reff{Figure}{fig:domain}. Moreover, $L_x$ and $L_y$ are the lengths of $\Omega$ in the respective horizontal axes, and $L_z$ is the lowest admissible value of the length of $\Omega$ in the vertical axis. We assume the width of $\Omega_1$ to be $10\%$ of $L_z$. 

\begin{figure}[H] 
\centering
\includegraphics[]{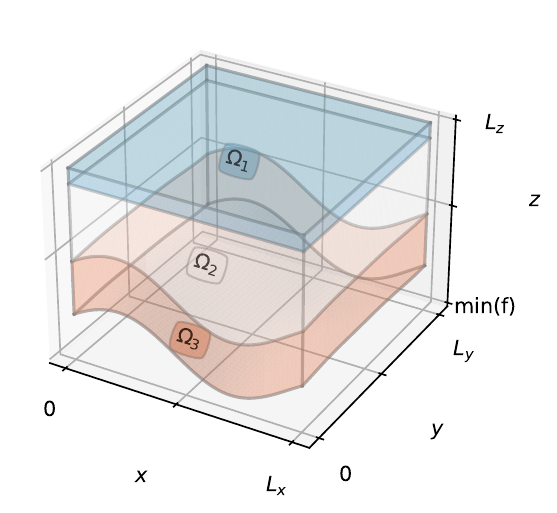} 
\caption{The domain, $\Omega$, of the proposed model.}
\label{fig:domain}
\end{figure}

\subsection{Pathology and dynamics}
\label{sec:pathology}

In this section, we describe the key features of the pathology regarding the proliferation of dysplastic cells, along with their interactions with epithelial cells, HPV viral particles and immune cells. We use $u$ to symbolise the concentration of epithelial cells, $v$ for the concentration of HPV viral particles, $c$ for the concentration of dysplastic cells and $w$ for the concentration of immune cells.

\subsubsection{Epithelial cells} 

We begin by assuming that epithelial cells proliferate in a constant manner. The proliferative part of the epithelium is the basal one \cite{kierszenbaum2015histology}. Therefore, we assume that the constant proliferation term of the epithelial cell population, $s_u$ to have the following form:

\begin{equation*}
    s_u = \begin{cases}
        0\;,       & \text{in $\Omega_1$} \\
        0\;,       & \text{in $\Omega_2$} \\
        s_{u_3}\;, & \text{in $\Omega_3$}\;.
    \end{cases}
\end{equation*}

Fick's law states that a substance's diffusive flux is proportional to the gradient of the substance's concentration. This means that a normally distributed substance would be diffused equally in every spatial direction. This doesn't seem to apply to epithelial cells, which are distributed across the vaginal epithelium. The superficial cells exfoliate continuously, while the basal cells move towards the surface and replace them \cite{kierszenbaum2015histology}. Therefore, we assume that a modified Fick's law governs the movement of epithelial cells, with their diffusive flux being equal to
\begin{equation*}
     \vb*{D_u} \odot \grad{u} \;,
\end{equation*}
with $\odot$ being the Hadamard product and 
\begin{equation*}
    \vb*{D_u} = \qty(D_{u_{xy}},D_{u_{xy}},D_{u_z}) 
\end{equation*}
being the collective diffusion coefficient of $u$, with $D_{u_{xy}} < D_{u_z}$. Therefore, the diffusion term for the concentration of epithelial cells is equal to $$\div[ \vb*{D_u} \odot \grad{u}] \;.$$ This term, instead of describing random motility like a diffusion described by Fick's law would, describes a type of weighted random motility, with a skew towards faster movement on the $z$-axis.

Additionally, we assume that epithelium cells decay exponentially, and at a faster rate when in $\Omega_1$, due to the lack of nutrients. Therefore, we have that the term $-m_u\qty(\vb*{x}) u$ models the exponential decay of epithelial cells, with the rate of exponential decay of $u$ being:

\begin{equation*}
    m_u = \begin{cases}
        m_{u_1}\;,       & \text{in $\Omega_1$} \\
        m_{u_2}\;,       & \text{in $\Omega_2$} \\
        m_{u_3}\;,       & \text{in $\Omega_3$}\;,
    \end{cases}
\end{equation*}
where $m_{u_1} < \min{\qty(m_{u_2},m_{u_3})}$\;.

We assume that virus particles mutate epithelial cells into dysplastic cells with a constant rate. We model this mutation with the term $-k v u$, with $k$ being the rate of mutation and  the negative sign indicating that epithelial cells are lost during this procedure.

Furthermore, we assume zero flux boundary conditions. Thus,
\begin{equation*}
    \pdv{u}{n} = 0 \;, \quad \text{on } \partial{\Omega} \;.
\end{equation*}

The initial condition of the epithelium cells, $u_\infty$, is to be extracted as follows. Using the finite element analysis software COMSOL Multiphysics$^\circledR$ 6.1 \cite{comsol}, we numerically solve, until a time point, the following problem:
\begin{subequations}
    \begin{align*}
            \pdv{u}{t} &= \div[ \vb*{D_u} \odot \grad{u}] + s_u - m_u u - k v u\;, &&\text{in $\Omega^\circ$} \\
            \pdv{u}{n} &= 0\;, &&\text{on $\partial\Omega$} \\
            u(\cdot,0) &= 0\;, &&\text{in $\Omega$} \;.
    \end{align*}
\end{subequations}
For each $f$, its solution seems to be converging towards an equilibrium, which we consider to be $u_\infty$. In this way, we initialize the distribution of epithelial cells across $\Omega$ before any pathological interaction takes place. 

Summing up, we have that
\begin{subequations}
    \begin{align*}
            \pdv{u}{t} &= \div[ \vb*{D_u} \odot \grad{u}] + s_u - m_u u - k v u\;, &&\text{in $\Omega^\circ$} \\
            \pdv{u}{n} &= 0\;, &&\text{on $\partial\Omega$} \\
            u(\cdot,0) &= u_\infty \;, &&\text{in $\Omega$} \;.
    \end{align*}
\end{subequations}


\subsubsection{Viral particles} 

HPV infection is initiated when the virus comes into contact with the basal epithelial cells, which are typically reached through a microabrasion in the epithelial tissue \cite{munger2004mechanisms,schiller2010current,pevsut2021human}. The initial contact between the epithelium and the viral particles happens at the superficial epithelium. Additionally, the frequency of sexual intercourse or other intimate skin-to-skin contact is a risk factor \cite{veldhuijzen2010factors}. Therefore, in order to capture the above key mechanisms, we assume that there's a spatially small, time-dependent constant source of HPV viral particles at the superficial epithelium. We write that source at point $(\vb*{x},t)\in\Omega\times\mathbb{R}_{\ge 0}$ as $s_v(\vb*{x},t)$. 

Consequently, we assume that the aforementioned movement of viral particles from the superficial epithelium to the basal layer happens in two ways. The first one is through random motility in space, so we model it using the diffusion term $D_v \Delta v$, with $D_v$ being the diffusion coefficient of $v$. The second one is through a directed movement towards the epithelial cells, therefore we model it using the advection term $- a_v \div[\tau_v\frac{v}{v+b} \grad{u}]$ , with $\tau_v$ being the velocity field coefficient of $v$ towards $u$, and $b$ being the $v$ value for half-maximal velocity of $v$ towards $u$. We choose the velocity field of the viral particles to be a Michaelis-Menten function.

Additionally, we assume that the HPV viral particles decay exponentially. Hence, we introduce the term $-m_v v$, with $m_v$ being the rate of exponential decay of $v$.

Furthermore, the number of HPV viral particles may increase in dysplastic epithelial cells due to the uncontrollable expression of HPV oncogenes, leading to the maintenance of the transformed phenotype and ultimately promoting the development of cancer \cite{pevsut2021human}. Thus, we assume that dysplastic cells increase the concentration of viral particles. To model this, we introduce the nonlinear logistic term
\begin{equation*}
    r_v v \qty(p c - \frac{v}{q}) \;,
\end{equation*}
where $r_v$ is the growth rate of $v$, $q$ is a carrying-capacity-like quantity for $v$, and $p$ is an inverse-carrying-capacity-like quantity for $v$. We note that $q$ and $p$ have the same and the inverse dimensions, respectively, as the carrying capacity of the common logistic term. For $c \neq 0 $, we notice that the above term is equal to $ r_v p c v \qty(1 - \dfrac{v}{pqc})$\;. In this form, it is clear that the carrying capacity and the growth rate of the concentration of HPV viral particles is proportional to the concentration of dysplastic cells. 

Finally, we assume zero flux boundary conditions and an initial condition equal to zero.

Summing up, we have that
\begin{subequations}
    \begin{align*}
            \pdv{v}{t} &= D_v \Delta v - a_v \div[\tau_v\frac{v}{v+b} \grad{u}] + s_v - m_v v + r_v v \qty(p c - \frac{v}{q})\;, &&\text{in $\Omega^\circ$} \\
            \pdv{v}{n} &= 0\;, &&\text{on $\partial\Omega$} \\
            v(\cdot,0) &= 0 \;, &&\text{in $\Omega$} \;.
    \end{align*}
\end{subequations}

\subsubsection{Dysplastic cells}

Regarding the movement of dysplastic cells, we employ the same idea as the one we used for the epithelial cells with a weighted random motility type of movement. Thus, we have that the diffusion term for the dysplastic cells is $$\div[ \vb*{D_c} \odot \grad{c}] \;, $$
with $\vb*{D_c}$ being the collective diffusion coefficient of $c$.

Furthermore, we assume that in $\Omega_2 \cup \Omega_3$, without immune response, neoplastic cells do not decay \cite{park2023apoptosis}. However, we assume that in $\Omega_1$ they decay exponentially, due to the lack of nutrients. Therefore, we have that the term $-m_c\qty(\vb*{x}) u$ models the exponential decay of epithelial cells, with 

\begin{equation*}
    m_c(\vb*{x}) = \begin{cases}
        m_{c_1}\;,       & \text{in $\Omega_1$} \\
        0\;,             & \text{otherwise}\;, \\

    \end{cases}
\end{equation*}
being the rate of exponential decay of $c$.

As we discussed for the case of HPV virus particles, we assume that the epithelial cells mutate into dysplastic cells with a rate of $k v u$. The term's positive sign indicates that the concentration of dysplastic cells increases during this procedure.

Moreover, we assume that immune cells kill dysplastic cells with a Holling type I functional response. Hence, we have that the term describing this procedure is $-dwc$\;, with $d$ being the rate of $w$-induced $c$ death.

Finally, we assume zero flux boundary conditions and an initial condition equal to zero.

Summing up, we have that
\begin{subequations}
    \begin{align*}
            \pdv{c}{t} &= \div[ \vb*{D_c} \odot \grad{c}] - m_c c + k v u - d w c\;, &&\text{in $\Omega^\circ$} \\
            \pdv{c}{n} &= 0\;, &&\text{on $\partial\Omega$} \\
            c(\cdot,0) &= 0 \;, &&\text{in $\Omega$} \;.
    \end{align*}
\end{subequations}

\subsubsection{Immune cells}

Different types of immune cells, such as leukocytes, neutrophils, and monocytes, preferentially migrate through areas of the basement membrane that have lower protein deposition, such as laminin and collagen IV. These cells can migrate through existing openings in the basement membrane. For instance, monocytes are highly deformable and can squeeze through these openings, while neutrophils can lead to remodeling and enlargement of these sites. Dendritic cells, another type of immune cell, have been shown to migrate through preexisting openings in the lymphatic basement membrane by widening these small gaps. After these gaps are widened by cells, they return to a baseline slightly larger than the original gap size, indicating the mechanical plasticity of the basement membrane \cite{chang2019beyond}. Therefore, we assume that the greater the concentration of dysplastic cells near the basement membrane, the higher the influx of immune cells that pass through it. Hence, we have that
\begin{equation*}
    \pdv{w}{n} = jc\;, \quad \text{on }\gr{f} \;,
\end{equation*}
where $j$ is the flux coefficient through the basement membrane, and $$\gr{f}=\set*{\qty(x,y,z) \in \qty[0,L_x]\cross\qty[0,L_y] \cross \qty[\min{f},L_z] | z = f\qty(x,y) }$$ stands for the graph of $f$. 

Regarding the movement of immune cells, we assume that immune cells move both in a random and in a directed way. The random movement of immune cells is modeled by the diffusion term $D_w \Delta w$\;, with $D_w$ being the diffusion coefficient of $w$. On the other hand, the directed movement of immune cells towards dysplastic cells is modeled by the advection term $-a_w \div[\tau_w w \grad{c}]$, with $a_w$ being the advection coefficient of $w$\;, and $\tau_w$ being the velocity field coefficient of $w$ towards $c$.

Moreover, we assume that the immune cells decay exponentially. Hence, we introduce the term $-m_w w$, with $m_w$ being the rate of exponential decay of $w$.

Finally, we have that the initial concentration of immune cells is equal to zero.

Summing up, we have that
\begin{subequations}
    \begin{align*}
            \pdv{w}{t} &= D_w \Delta w - a_w \div[\tau_w w \grad{c}] - m_w w\;, &&\text{in $\Omega^\circ$} \\
            \pdv{w}{n} &= jc\;, &&\text{on } \gr{f} \; \\
            \pdv{w}{n} &= 0\;, && \text{on } \partial{\Omega} \setminus \gr{f} \\
            w(\cdot,0) &= 0 \;, &&\text{in $\Omega$} \;.
    \end{align*}
\end{subequations}


\subsection{The integrated model}
\label{model}

Integrating the histological and the pathological part of the model presented in \hyperref[sec:histology]{\S \ref*{sec:histology}} and \hyperref[sec:pathology]{\S \ref*{sec:pathology}}, respectively, we arrive at the IBVP that describes the problem in question.

Our model consists of the following system of nonlinear reaction-diffusion-advection equations

\begin{subequations} \label{eq:main_system}
    \begin{align}
        \pdv{u}{t} &= \div[ \vb*{D_u} \odot \grad{u}] + s_u - m_u u - k v u \label{eq:u}\\ 
        \pdv{v}{t} &= D_v \Delta v - a_v \div[\tau_v\frac{v}{v+b} \grad{u}] + s_v - m_v v + r_v v \qty(p c - \frac{v}{q}) \label{eq:v}\\
        \pdv{c}{t} &= \div[ \vb*{D_c} \odot \grad{c}] - m_c c + k v u - d w c \label{eq:c}\\
        \pdv{w}{t} &= D_w \Delta w - a_w \div[\tau_w w \grad{c}] - m_w w \label{eq:w}\;,
    \end{align}
\end{subequations}

in $\Omega^\circ$, with initial conditions

\begin{subequations} 
    \begin{align*}
        u\qty(\cdot,0) &= u_{\infty}\\ 
        v\qty(\cdot,0) &= 0 \\
        c\qty(\cdot,0) &= 0 \\
        w\qty(\cdot,0) &= 0 \;,
    \end{align*}
\end{subequations}

in $\Omega$, along with boundary conditions

\begin{subequations} 
    \begin{align*}
        \pdv{u}{n} &= \pdv{v}{n} = \pdv{c}{n} = 0\;, && \text{on } \partial{\Omega}\\
        \pdv{w}{n} &= 0\;, && \text{on } \partial{\Omega} \setminus \gr{f} \\
        \pdv{w}{n} &= jc\;, && \text{on } \gr{f} \;.
    \end{align*}
\end{subequations}

\reff{Table}{tab:term_discription} and \reff{Table}{tab:param_discription} list the terms, as well as the variables and parameters, respectively, of our model, along with a brief description. In \reff{Table}{tab:param_discription3} we present scales of the variables and estimations of the parameters, which will be used later on during the numerical simulations. 
 
\begin{table}
    \centering
        \begin{NiceTabular}{ccc}[]
            \toprule
            \textbf{Deriv.} & \textbf{Term}     & \textbf{Description} \\
            \midrule
            \Block{4-1}{$\dfrac{\partial u}{\partial t}$}       & $\div[\vb*{D_u}\odot\grad{u}]$    & Weighted random motility of epithelial cells \\
                                            & $s_u$             & Constant source of epithelial cells \\
                                            & $-m_u u$          & Exponential decay of epithelial cells    \\
                                            & $-kvu$            & Mutation of epithelial cells to dysplastic cells due to viral particles \\
            \midrule
            \Block{5-1}{$\dfrac{\partial v}{\partial t}$}       & $D_v \Delta v$    & Random motility of viral particles \\
                                            & $- a_v \div[\tau_v\frac{v}{v+b} \grad{u}]$ & Directed movement of viral particles towards epithelial cells\\
                                            & $s_v$             & Constant source of viral particles  \\
                                            & $-m_v v$          & Exponential decay of viral particles  \\
                                            & $r_v v \qty(p c - \frac{v}{q})$    & Logistic growth of viral particles  \\
            \midrule
            \Block{4-1}{$\dfrac{\partial c}{\partial t}$}       & $\div[\vb*{D_c}\odot\grad{c}]$& Weighted random motility of dysplastic cells  \\
                                            & $-m_c c$          & Exponential decay of dysplastic cells \\
                                            & $kvu$             & Mutation of epithelial cells to dysplastic cells due to viral particles \\
                                            & $-dwc$            & Immune-cell-induced dysplastic cell death  \\            
            \midrule
            \Block{3-1}{$\dfrac{\partial w}{\partial t}$}       & $D_w \Delta w$    & Random motility of immune cells  \\
                                            & $- a_w \div[\tau_w w \grad{c}]$ & Directed movement of immune cells towards dysplastic cells  \\
                                            & $-m_w w$          & Exponential decay of immune cells  \\ 
            \bottomrule
        \end{NiceTabular}
    \caption{Description of the terms of the model.}
    \label{tab:term_discription}
\end{table}

\begin{table}
    \centering
        \begin{NiceTabular}{cccc}[]
            \addlinespace[-\aboverulesep] \cmidrule[\heavyrulewidth]{2-4}
                                            & \textbf{Name} & \textbf{Description} & \Block{1-1}{\textbf{Physical}\\ \textbf{dimension}} \\
            \cmidrule{2-4}
            \Block{4-1}{\rotate Independent \\ variables}   & $x$ & Length & L\\
                                                            & $y$ & Length & L\\
                                                            & $z$ & Length & L\\
                                                            & $t$ & Time   & T\\
            \cmidrule{2-4}                                
            \Block{4-1}{\rotate Dependent \\ variables}  & $u$ & Concentration of epithelial cells & $\#\,\text{L}^{-3}$\\
                                            & $v$ & Concentration of viral particles & $\#\,\text{L}^{-3}$\\
                                            & $c$ & Concentration of dysplastic cells& $\#\,\text{L}^{-3}$\\
                                            & $w$ & Concentration of immune cells& $\#\,\text{L}^{-3}$\\
            \cmidrule{2-4}
            \Block{27-1}{\rotate Parameters} & $\vb*{D_u}$ & Collective diffusion coefficient of $u$ & $\text{L}^{2}\,\text{T}^{-1}$\\
                                             & $s_u$ & Constant source of $u$& $\#\,\text{L}^{-3}\,\text{T}^{-1}$ \\
                                             & $m_u$ & Rate of exponential decay of $u$& $\text{T}^{-1}$\\
                                             & $k$ & Rate of mutation of $u$ to $c$& $\#^{-1}\,\text{L}^{3}\,\text{T}^{-1}$\\
                                             & $D_v$ & Diffusion coefficient of $v$& $\text{L}^{2}\,\text{T}^{-1}$\\
                                             & $a_v$ & Advection coefficient of $v$& $\text{L}^{2}$\\
                                             & $\tau_v$ & Velocity field coefficient of $v$ towards $u$& $\text{T}^{-1}$\\
                                             & $b$ & $v$ value for half-maximal velocity of $v$ towards $u$& $\#\,\text{L}^{-3}$\\
                                             & $s_v$ & Constant source of $v$& $\#\,\text{L}^{-3}\,\text{T}^{-1}$\\
                                             & $m_v$ & Rate of exponential decay of $v$& $\text{T}^{-1}$\\
                                             & $r_v$ & Growth rate of $v$ & $\text{T}^{-1}$\\
                                             & $p$ & Inverse-carrying-capacity-like quantity of $v$ & $\#^{-1}\,\text{L}^{3}$\\
                                             & $q$ & Carrying-capacity-like quantity of $v$ & $\#\,\text{L}^{-3}$\\
                                             & $\vb*{D_c}$ & Collective diffusion coefficient of $c$& $\text{L}^{2}\,\text{T}^{-1}$\\
                                             & $m_c$ & Rate of exponential decay of $c$& $\text{T}^{-1}$\\
                                             & $d$ & Rate of $w$-induced $c$ death& $\#^{-1}\,\text{L}^{3}\,\text{T}^{-1}$\\
                                             & $D_w$ & Diffusion coefficient of $w$& $\text{L}^{2}\,\text{T}^{-1}$\\
                                             & $a_w$ & Advection coefficient of $w$& $\text{L}^{2}$\\
                                             & $\tau_w$ & Velocity field coefficient of $w$ towards $c$& $\#^{-1}\,\text{L}^{3}\,\text{T}^{-1}$\\
                                             & $m_w$ & Rate of exponential decay of $w$& $\text{T}^{-1}$\\
                                             & $j$ & Flux coefficient through the basement membrane& $\text{L}^{-1}$\\
                                             & $L_x$ & Length of $\Omega$ in $x$-axis& $\text{L}$\\
                                             & $L_y$ & Length of $\Omega$ in $y$-axis& $\text{L}$\\
                                             & $L_z$ & Lowest admissible value of the length of $\Omega$ in $z$-axis& $\text{L}$\\
                                             & $\ell$ & Amplitude coefficient & $\text{L}$\\
                                             & $\phi$ & Phase angle & $\text{L}$\\
                                             & $\beta$ & Sprectral exponent & --\\
                                             & $h$ & Width of the basal layer& $\text{L}$\\
            \cmidrule[\heavyrulewidth]{2-4} \addlinespace[-\belowrulesep]
        \end{NiceTabular}
    \caption{Description of the independent and dependent variables as well as parameters of the model, along with their units.}
    \label{tab:param_discription}
\end{table}

\begin{table}
    \centering
        \begin{NiceTabular}{cccc}[]
            \addlinespace[-\aboverulesep] \cmidrule[\heavyrulewidth]{2-4}
                                            & \textbf{Name} & \textbf{Value} & \textbf{Unit} \\
            \cmidrule{2-4}
            \Block{8-1}{\rotate Scales }    & $x_0$ & $1.00 \cdot 10^{2}$ & $\mu \text{m}$\\
                                            & $y_0$ & $1.00 \cdot 10^{2}$ & $\mu \text{m}$\\
                                            & $z_0$ & $1.00 \cdot 10^{2}$ & $\mu \text{m}$\\
                                            & $t_0$ & 1.00 & d\\                      
                                            & $u_0$ & $2.64$ & $\#\,\mu \text{m}^{-3}$\\
                                            & $v_0$ & $2.70 \cdot 10^{1}$ & $\#\,\mu \text{m}^{-3}$\\
                                            & $c_0$ & $2.70 \cdot 10^{1}$ & $\#\,\mu \text{m}^{-3}$\\
                                            & $w_0$ & $4.30 \cdot 10^{-2}$ & $\#\,\mu \text{m}^{-3}$\\
            \cmidrule{2-4}
            \Block{29-1}{\rotate Parameters} & $\vb*{D_u}$ & $\qty(8.16\cdot 10^{2},8.16\cdot 10^{2},2.94\cdot 10^{4})$ & $\mu \text{m}^{2}\,\text{d}^{-1}$\\
                                             & $s_{u_3}$ & $2.10\cdot 10^{-1}$ & $\#\,\mu \text{m}^{-3}\,\text{d}^{-1}$ \\
                                             & $m_{u_1}$ & $8.00\cdot 10^{-1}$ & $\text{d}^{-1}$\\
                                             & $m_{u_2}$ & $5.00\cdot 10^{-2}$ & $\text{d}^{-1}$\\
                                             & $m_{u_3}$ & $5.00\cdot 10^{-2}$ & $\text{d}^{-1}$\\
                                             & $k$   & $1.40 \cdot 10^{1}$ & $\#^{-1}\,\mu \text{m}^{3}\,\text{d}^{-1}$\\
                                             & $D_v$ & $4.90 \cdot 10^{4}$ & $\mu \text{m}^{2}\,\text{d}^{-1}$\\
                                             & $a_v$ & $1.04 \cdot 10^{4}$ & $\mu \text{m}^{2}$\\
                                             & $\tau_v$ & 1.00 & $\text{d}^{-1}$\\
                                             & $b$   & $2.10 \cdot 10^{-2}$ & $\#\,\mu \text{m}^{-3}$\\
                                             & $s_{v}$ & varied & $\#\,\mu \text{m}^{-3}\,\text{d}^{-1}$\\
                                             & $m_v$ & $9.00 \cdot 10^{-4}$ & $\text{d}^{-1}$\\
                                             & $r_v$ & $1.00 \cdot 10^{-3}$ & $\text{d}^{-1}$\\
                                             & $p$ & $2.33 \cdot 10^{1}$ & $\#^{-1}\,\mu \text{m}^{3}$\\
                                             & $q$ & $0.43 \cdot 10^{-1}$ & $\#^{-1}\,\mu \text{m}^{-3}$\\                   
                                             & $\vb*{D_c}$ & $\qty(8.16\cdot 10^{2}, 8.16\cdot 10^{2}, 1.22\cdot 10^{5})$ & $\mu \text{m}^{2}\,\text{d}^{-1}$\\
                                             & $m_{c_1}$ & $8.00\cdot 10^{-1}$ & $\text{d}^{-1}$\\
                                             & $d$ & $2.33 \cdot 10^{1}$ & $\#^{-1}\,\mu \text{m}^{3}\,\text{d}^{-1}$\\
                                             & $D_w$    & $4.90 \cdot 10^{4}$ & $\mu \text{m}^{2}\,\text{d}^{-1}$\\
                                             & $a_w$    & $8.20 \cdot 10^{-2}$ & $\mu \text{m}^{2}$\\
                                             & $\tau_w$ & $2.33 \cdot 10^{1}$ & $\#^{-1}\,\mu \text{m}^{3}\,\text{d}^{-1}$\\
                                             & $m_w$ & $1.00\cdot 10^{-2}$ & $\text{d}^{-1}$\\
                                             & $j$ & varied & $\mu \text{m}^{-1}$\\
                                             & $L_x$ & $1.00 \cdot 10^{2}$ & $\mu \text{m}$\\
                                             & $L_y$ & $1.00 \cdot 10^{2}$ & $\mu \text{m}$\\
                                             & $L_z$ & $1.00 \cdot 10^{2}$ & $\mu \text{m}$\\
                                             & $\ell$ & $\mathcal{N}_2\qty( \vb{0}_2, \vb*{I}_2)$ & $\mu \text{m}$\\
                                             & $\phi$ & $\mathcal{U}\qty( \qty[-\frac{\pi}{2},\frac{\pi}{2}]^2)$ & $\mu \text{m}$\\
                                             & $\beta$ & 6.50 & --\\
                                             & $h$ & $3.00 \cdot 10^{1}$ & $\mu \text{m}$\\
            \cmidrule[\heavyrulewidth]{2-4} \addlinespace[-\belowrulesep]
        \end{NiceTabular}
    \caption{Scales of the variables and estimated parameter values of the model. The value of the ratio $\frac{h}{L_z}\approx 30\%$ is taken from \cite{chandrasoma2017gerd}, whereas the value of $L_z$ is taken from \cite{mescher2018junqueira}. The rest of the parameters are estimated.}
    \label{tab:param_discription3}
\end{table}

Finally, we note that by utilizing common techniques of non-dimensionalisation, we can eliminate only 3 of the 32 parameters in total. 

\section{Dysplasia progression: three key factors} 
\label{scenarios}

In the present section, we numerically solve the proposed problem, in order to highlight the importance of the shape of the basement membrane, the strength of the immune response, and the frequency of viral exposure. All numerical simulations are carried out using the finite element analysis software COMSOL Multiphysics$^\circledR$ 6.1.

The values of the parameters are as in \reff{Table}{tab:param_discription3}, unless otherwise noted. In order to incorporate the effect of the exposure to HPV we assume that $s_v\neq 0$, in particular 
\begin{equation*}
    s_v = \begin{cases}
        6.40 \cdot 10^{-2}\; \mu\text{m}^{3}\,\text{d}^{-1}\;, &  \text{in }\Omega_{1_s} \cross T_s \\
        0\;,       &  \text{otherwise} \;,    
    \end{cases}
\end{equation*}
for $$\Omega_{1_s}= {\left[4.50 \cdot 10^1, 5.50 \cdot 10^1 \right]}^2\times\left[9.00 \cdot 10^1 ,1.00 \cdot 10^2\right] \; \mu \text{m}^3\subsetneq \Omega_{1}$$ and for some $T_s \subsetneq \mathbb{R}_{\ge 0}$ which is left to be chosen. $T_s$ is the union of half-hour intervals, with each one representing a sexual encounter with an HPV positive partner \cite{blair2014can}. 

\subsection{Basement membrane's shape}
\label{s1}

Here, assuming that $T_s=[2,2 + \frac{1}{48}]\;\text{d}$ and $j=0$, we only vary the geometrical characteristics of the model. Therefore, drawing samples from $\ell$ and $\phi$, while keeping the rest of the parameters fixed, we arrive at \reff{Figure}{fig:sim_domain}. We notice that in some cases dysplatic cells vanish, whereas in others they persist. Hence, the ability of dysplatic cells to establish themselves is strongly dependent on the shape of the basement membrane. 

\begin{figure}[!ht]
    \begin{subfigure}[t]{1\textwidth}
        \centering
        \includegraphics[height = 1.5cm]{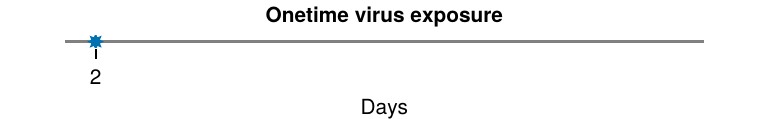}
        \setlength{\belowcaptionskip}{0cm}
        \caption*{}
        \label{fig:onetime5}
    \end{subfigure}
    \begin{subfigure}[t]{0.3\textwidth}
        \centering
        \includegraphics[width = 2in]{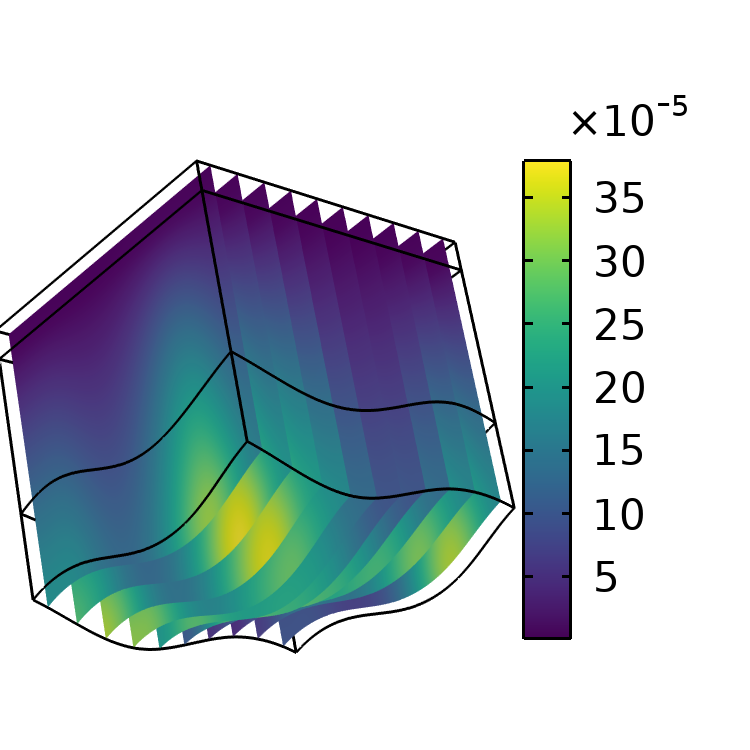}
        \caption{}
        \label{fig:sim_domain_a}
    \end{subfigure}
    \hspace*{\fill}%
    \begin{subfigure}[t]{0.3\textwidth}
        \centering
        \includegraphics[width = 2in]{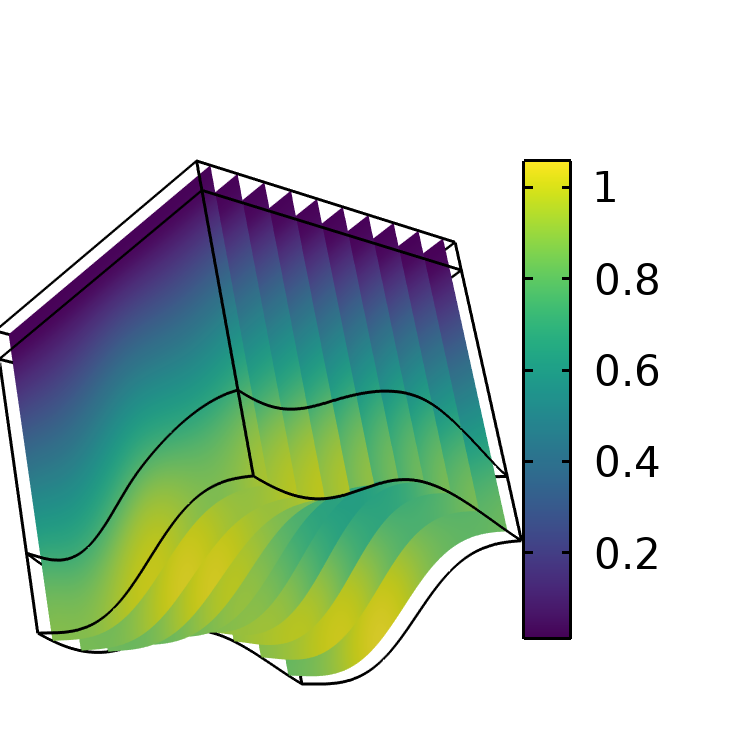}
        \caption{}
        \label{fig:sim_domain_b}
    \end{subfigure}
    \hspace*{\fill}%
    \begin{subfigure}[t]{0.3\textwidth}
        \centering
        \includegraphics[width = 2in]{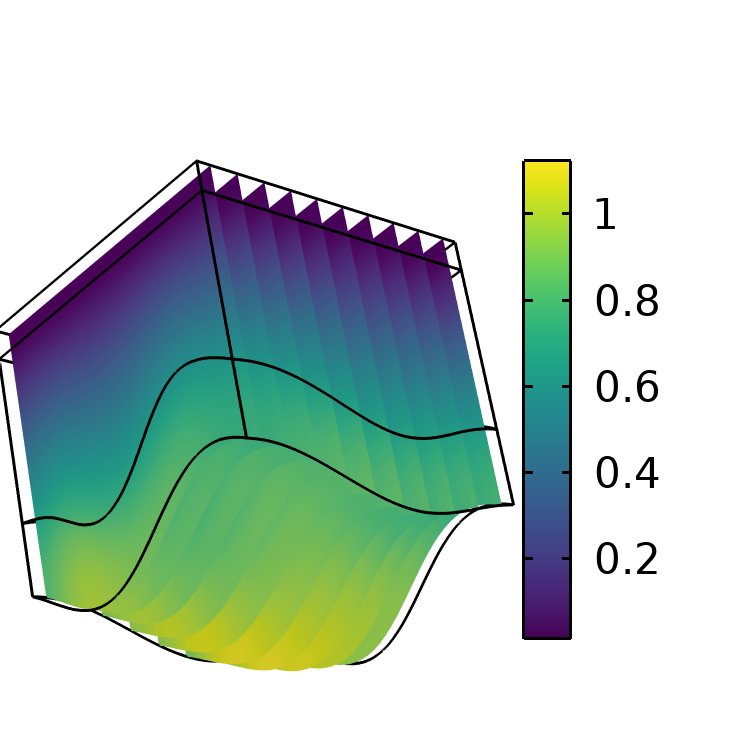}
        \caption{}
        \label{fig:sim_domain_c}
    \end{subfigure}
    \hspace*{\fill}%
    \begin{subfigure}[t]{0.3\textwidth}
        \centering
        \includegraphics[width = 2in]{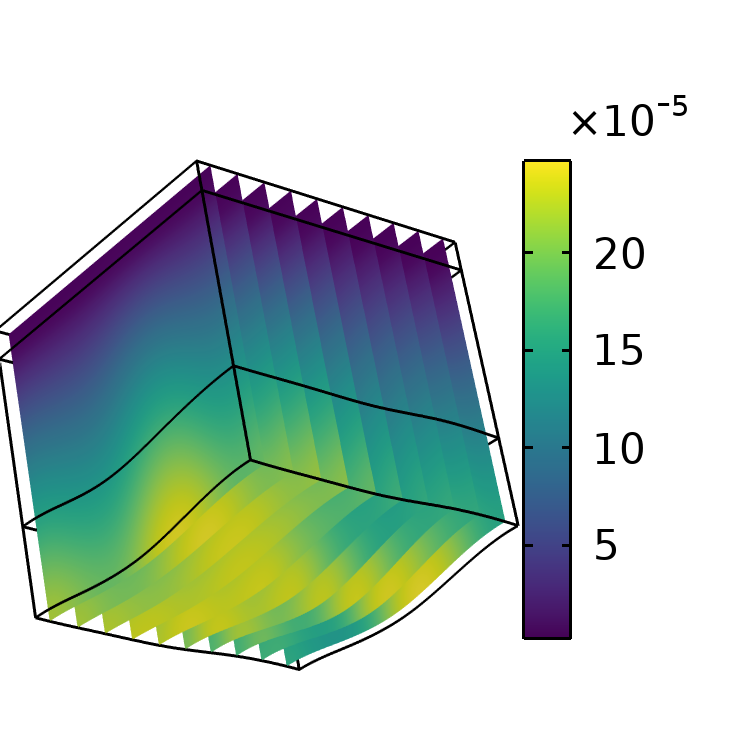}
        \caption{}
        \label{fig:sim_domain_d}
    \end{subfigure}
    \hspace*{\fill}%
    \begin{subfigure}[t]{0.3\textwidth}
        \centering
        \includegraphics[width = 2in]{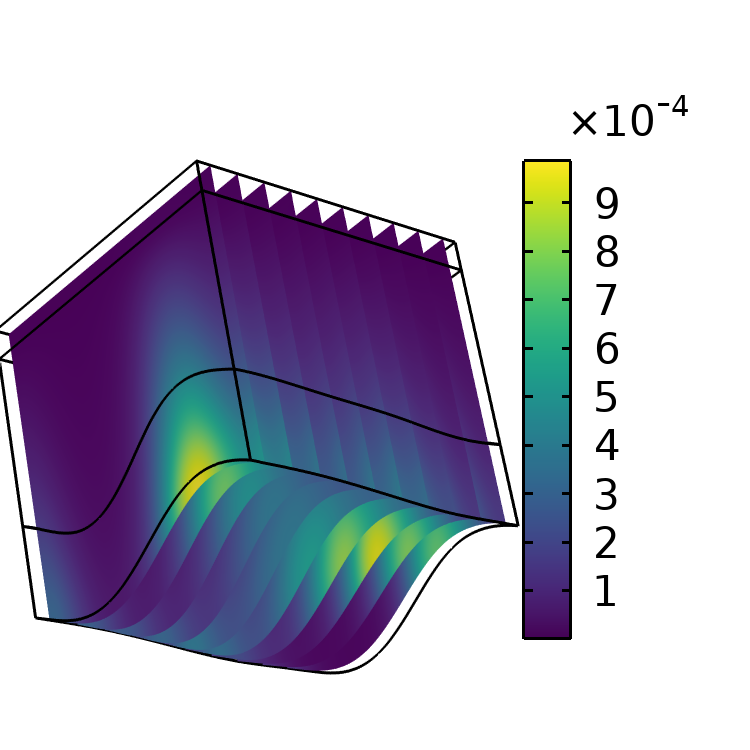}
        \caption{}
        \label{fig:sim_domain_e}
    \end{subfigure}
    \hspace*{\fill}%
    \begin{subfigure}[t]{0.3\textwidth}
        \centering
        \includegraphics[width = 2in]{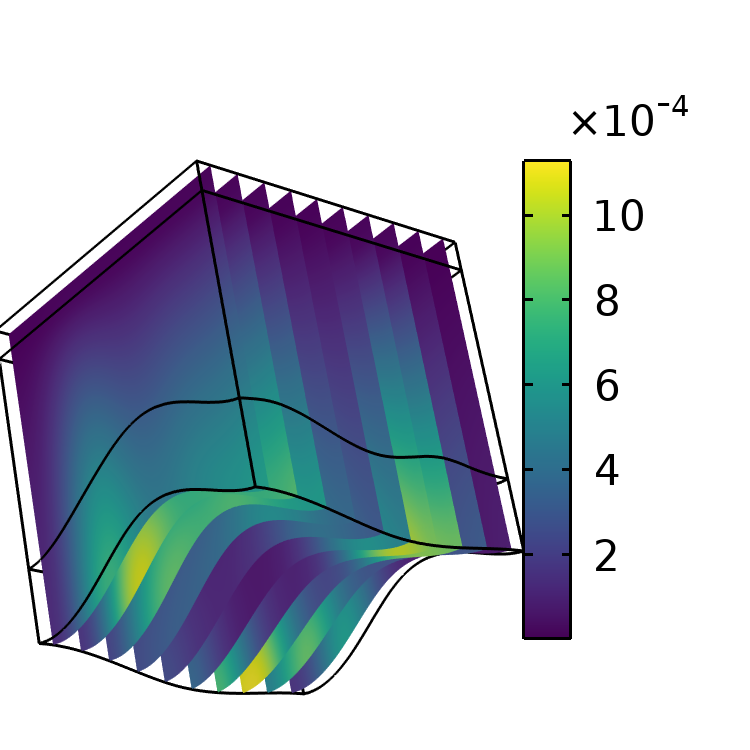}
        \caption{}
        \label{fig:sim_domain_f}
    \end{subfigure}
    \caption{Dysplastic cells after 3600 days. The geometry of the basement membrane varies between each simulation, while keeping the rest of the parameters fixed. We observe that this factor alone shapes the outcome of the infection. Namely, in simulations (a) and (d-f) the dysplastic cells tend to vanish, while in (b) and (c) they are established.}
    \label{fig:sim_domain}
\end{figure}

\subsection{Immune response's strength}
\label{s2}

Assuming that $T_s=[2,2 + \frac{1}{48}]\;\text{d}$ and $f$ is fixed, while only varying the value of $j$, we arrive at \reff{Figure}{fig:flux}. We deduce that as $j$ decreases, the establishment of dysplastic cells is facilitated. 

\begin{figure}[!ht]
    \begin{subfigure}[t]{1\textwidth}
        \centering
        \includegraphics[height = 1.5cm]{onetime.pdf}
        \setlength{\belowcaptionskip}{0cm}
        \caption*{}
    \end{subfigure}
    \begin{subfigure}[t]{0.3\textwidth}
        \centering
        \includegraphics[width = 2in]{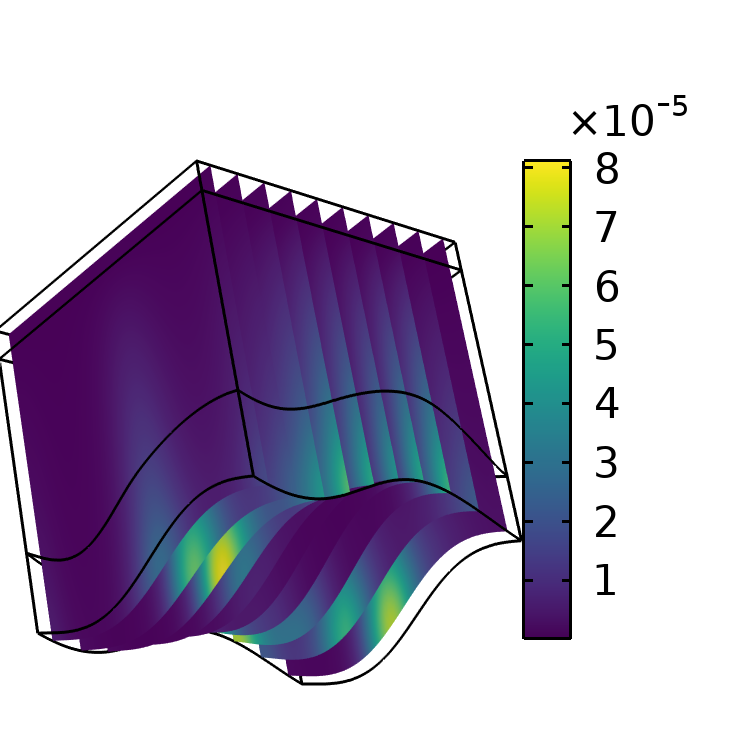}
        \caption{$j=3.86 \cdot 10^{3} \; \mu \text{m}^{-1}$}
        \label{fig:sim_flux_f}
    \end{subfigure}
    \hspace*{\fill}%
    \begin{subfigure}[t]{0.3\textwidth}
        \centering
        \includegraphics[width = 2in]{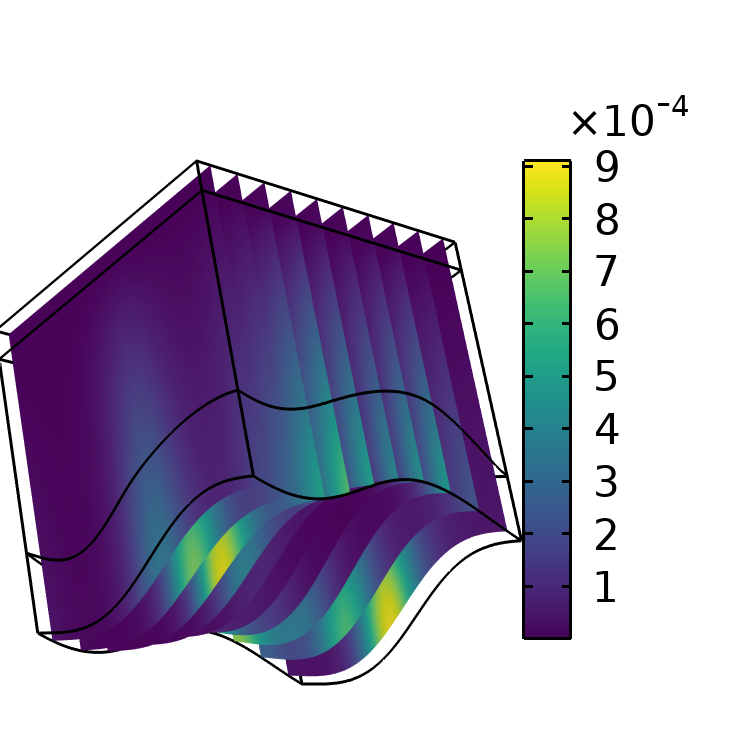}
        \caption{$j=3.86 \cdot 10^{2} \; \mu \text{m}^{-1}$}
        \label{fig:sim_flux_e}
    \end{subfigure}    
    \hspace*{\fill}%
    \begin{subfigure}[t]{0.3\textwidth}
        \centering
        \includegraphics[width = 2in]{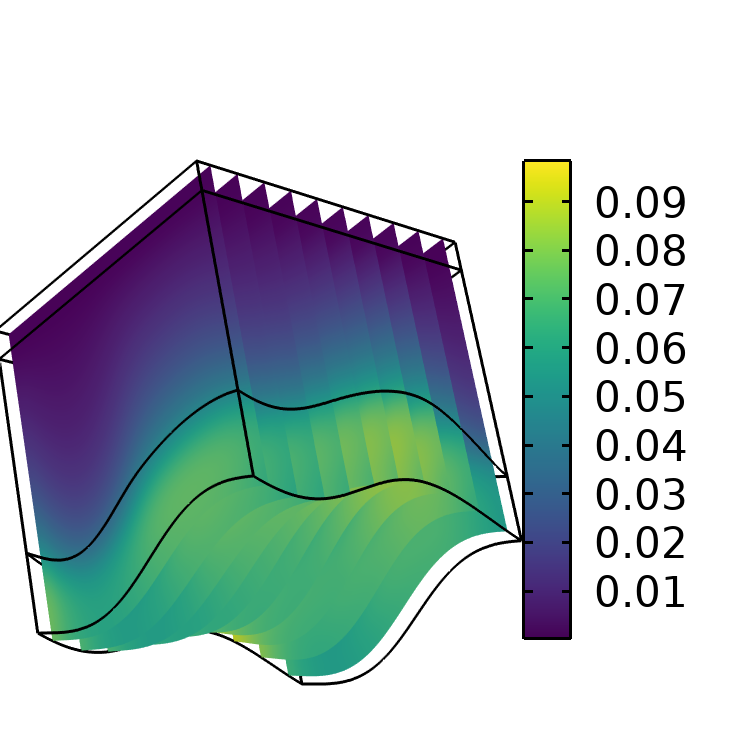}
        \caption{$j=3.86 \cdot 10^{1} \; \mu \text{m}^{-1}$}
        \label{fig:sim_flux_d}
    \end{subfigure}
    \hspace*{\fill}%
    \begin{subfigure}[t]{0.3\textwidth}
        \centering
        \includegraphics[width = 2in]{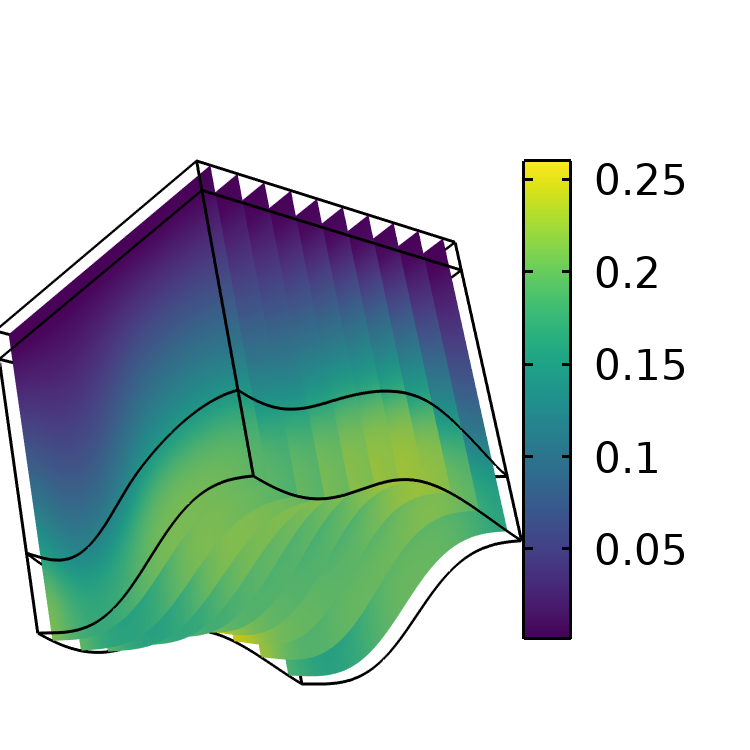}
        \caption{$j=3.86 \; \mu \text{m}^{-1}$}
        \label{fig:sim_flux_c}
    \end{subfigure}
    \hspace*{\fill}%
    \begin{subfigure}[t]{0.3\textwidth}
        \centering
        \includegraphics[width = 2in]{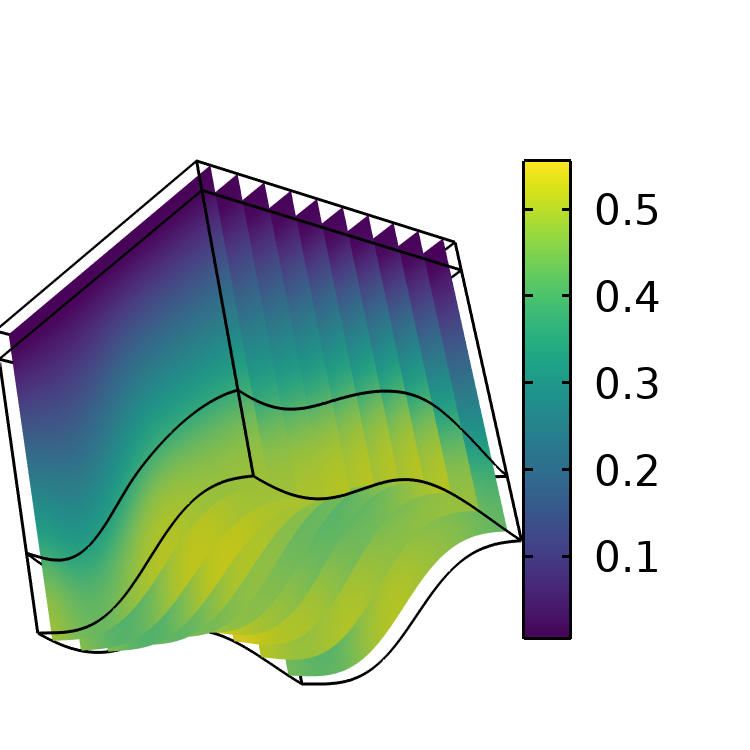}
        \caption{$j=3.86\cdot 10^{-1} \; \mu \text{m}^{-1}$}
        \label{fig:sim_flux_b}
    \end{subfigure}
    \hspace*{\fill}%
    \begin{subfigure}[t]{0.3\textwidth}
        \centering
        \includegraphics[width = 2in]{sim_domain_b_seed19.png}
        \caption{$j=0$}
        \label{fig:sim_flux_a}
    \end{subfigure}
    \caption{Dysplastic cells after 3600 days. $j$ varies between each simulation, while keeping the rest of the parameters fixed. We observe the qualitative change of the outcome as $j$ varies between $3.86 \cdot 10^{2}$ and $3.86 \cdot 10^{1} \; \mu \text{m}^{-1}$.}
    \label{fig:flux}
\end{figure}


\subsection{Viral exposure's frequency}
\label{s3}

Assuming that $j=3.86 \cdot 10^{1} \; \mu \text{m}^{-1}$ and $f$ is fixed, while only varying $T_s$, we arrive at \reff{Figure}{fig:times1}, \reff{Figure}{fig:times2} and \reff{Figure}{fig:times4}. We observe that the frequency of viral exposure determines the presence of dysplastic cells and increases the rate of their establishment. Therefore, this frequency has both qualitative and quantitative effect on the ultimate presence of precancerous lesions.

\begin{figure}[!ht]
    \begin{subfigure}[t]{1\textwidth}
        \centering
        \includegraphics[height = 1.5cm]{onetime.pdf}
        \setlength{\belowcaptionskip}{0cm}
        \caption*{}
        \label{fig:onetime7}
    \end{subfigure}
    \begin{subfigure}[t]{0.3\textwidth}
        \centering
        \includegraphics[width = 2in]{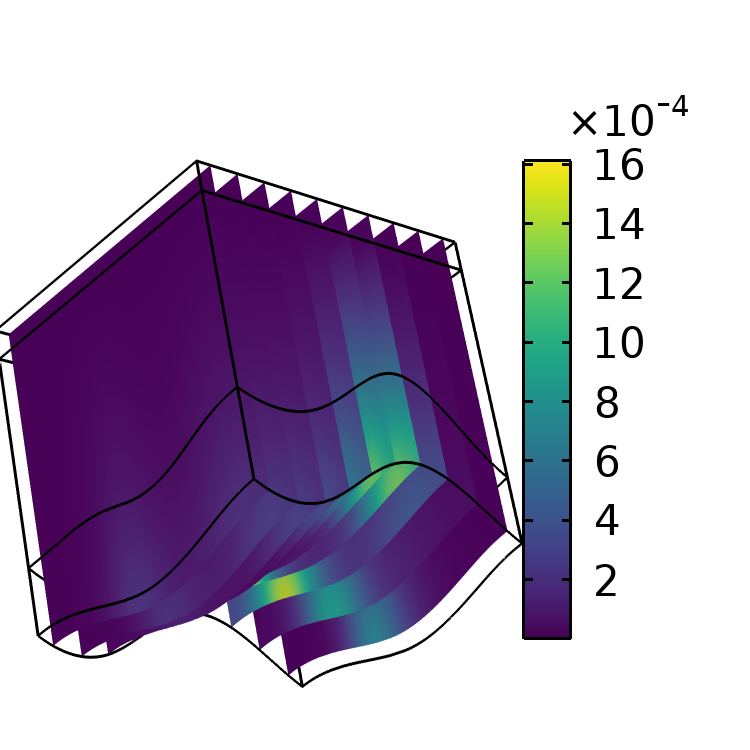}
        \caption{$t=100 \; \text{d}$}
        \label{fig:sim_times_1_100_a}
    \end{subfigure}
    \hspace*{\fill}%
    \begin{subfigure}[t]{0.3\textwidth}
        \centering
        \includegraphics[width = 2in]{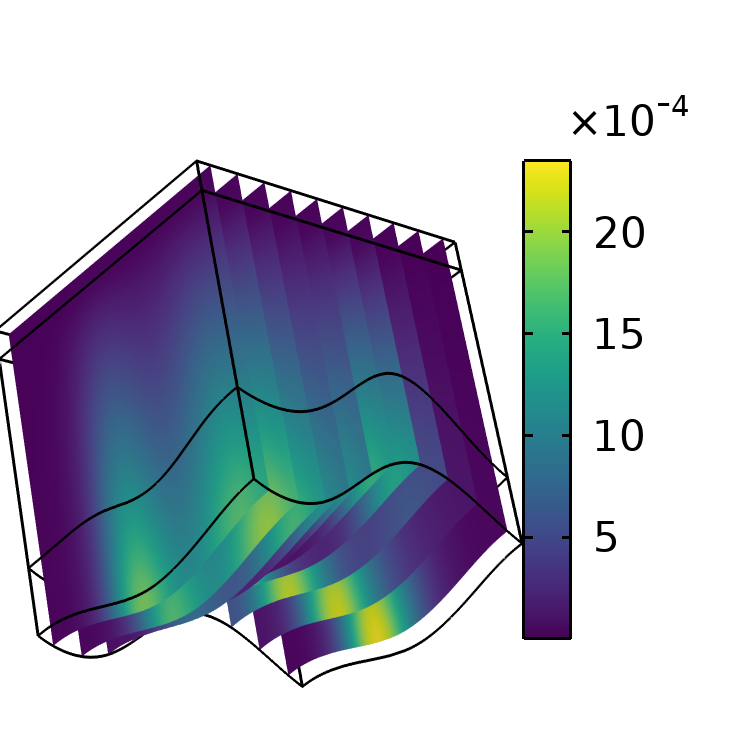}
        \caption{$t=800 \; \text{d}$}
        \label{fig:sim_times_1_800_b}
    \end{subfigure}
    \hspace*{\fill}%
    \begin{subfigure}[t]{0.3\textwidth}
        \centering
        \includegraphics[width = 2in]{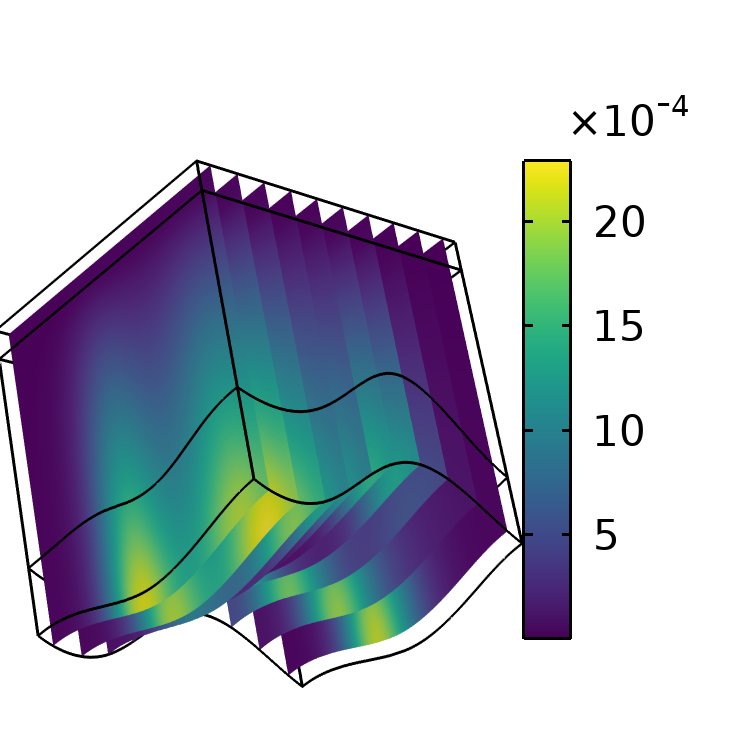}
        \caption{$t=1350 \; \text{d}$}
        \label{fig:sim_times_1_1350c}
    \end{subfigure}
    \caption{Snapshots of dysplastic cells for three different days. Here $T_s = [2,2 + \frac{1}{48}]\;\text{d}$. We observe that a onetime virus exposure is unable to cause dysplastic cell establishment, for this particular parameter set.}
    \label{fig:times1}
\end{figure}

\begin{figure}[!ht]
    \begin{subfigure}[t]{1\textwidth}
        \centering
        \includegraphics[height = 1.5cm]{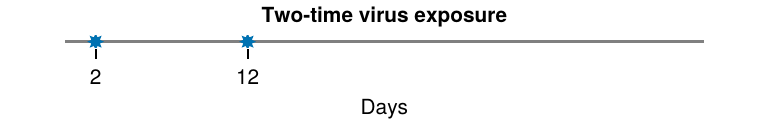}
        \setlength{\belowcaptionskip}{0cm}
        \caption*{}
        \label{fig:twotimes8}
    \end{subfigure}
    \begin{subfigure}[t]{0.3\textwidth}
        \centering
        \includegraphics[width = 2in]{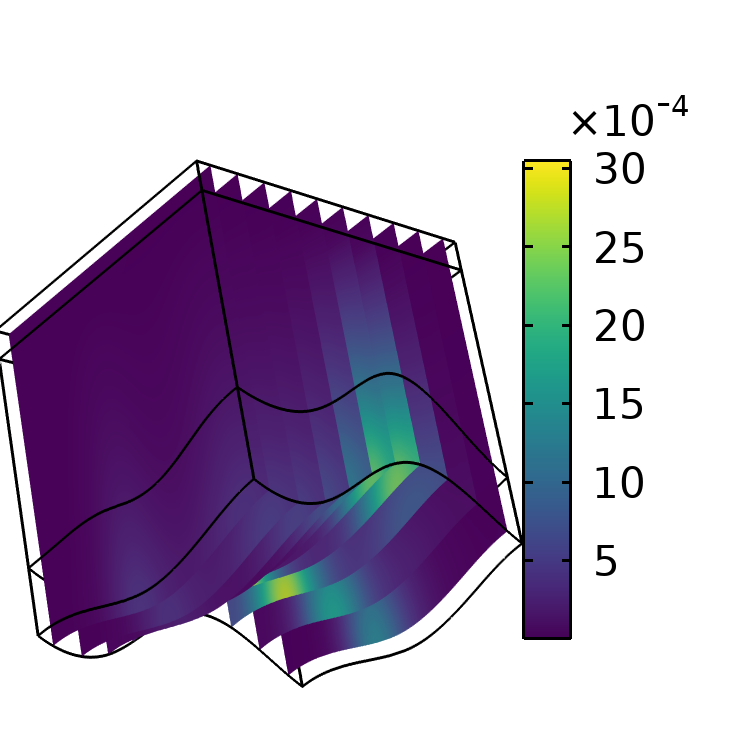}
        \caption{$t=100 \; \text{d}$}
        \label{fig:sim_times_2_100_a}
    \end{subfigure}
    \hspace*{\fill}%
    \begin{subfigure}[t]{0.3\textwidth}
        \centering
        \includegraphics[width = 2in]{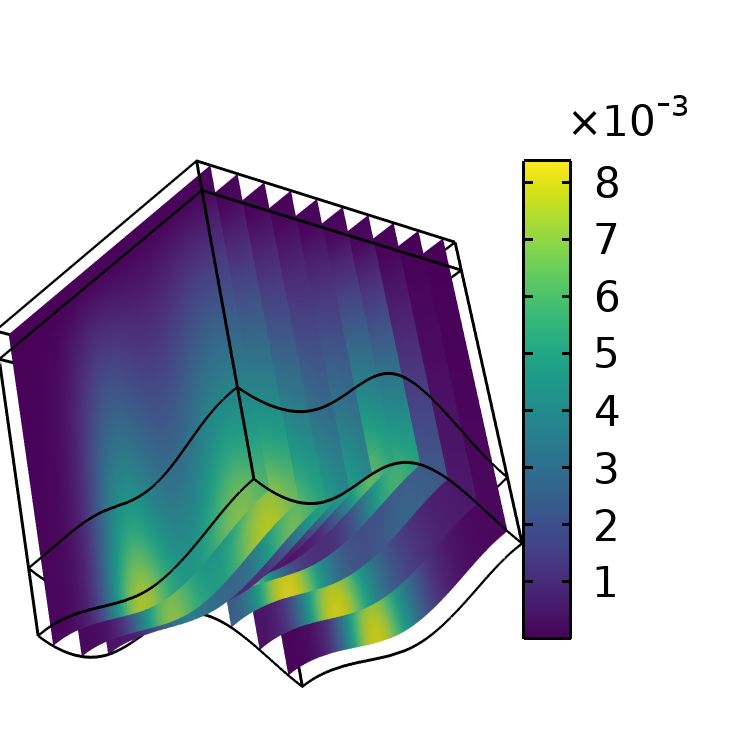}
        \caption{$t=800 \; \text{d}$}
        \label{fig:sim_times_2_800_b}
    \end{subfigure}
    \hspace*{\fill}%
    \begin{subfigure}[t]{0.3\textwidth}
        \centering
        \includegraphics[width = 2in]{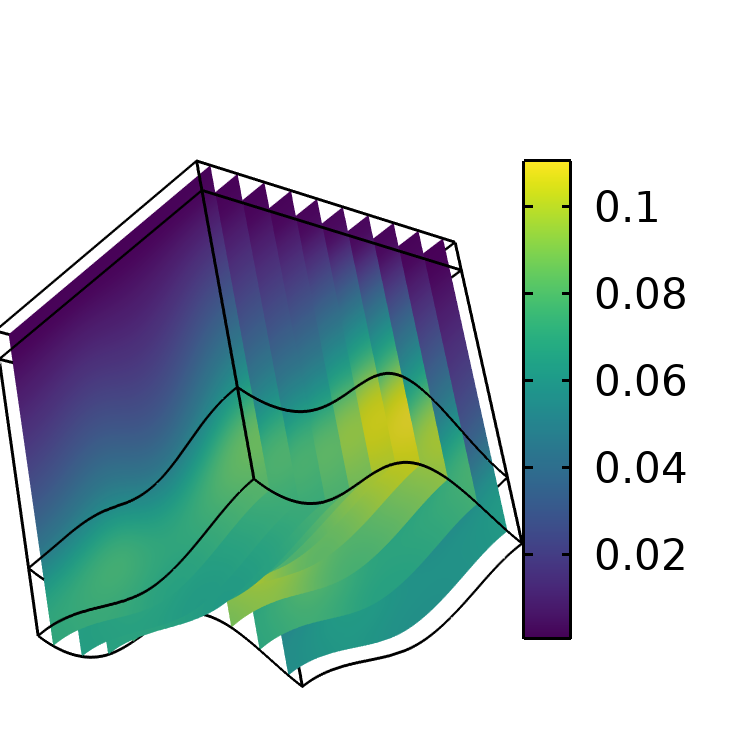}
        \caption{$t=1350 \; \text{d}$}
        \label{fig:sim_times_2_1350_c}
    \end{subfigure}
    \caption{Snapshots of dysplastic cells for three different days. Here $T_s = [2,2 + \frac{1}{48}]\cup[12,12 + \frac{1}{48}]\;\text{d}$. We observe that, contrary to a onetime virus exposure, a two-time exposure is able to cause dysplastic cell establishment, for the same parameter set.}
    \label{fig:times2}
\end{figure}

\begin{figure}[!ht]
    \begin{subfigure}[t]{1\textwidth}
        \centering
        \includegraphics[height = 1.5cm]{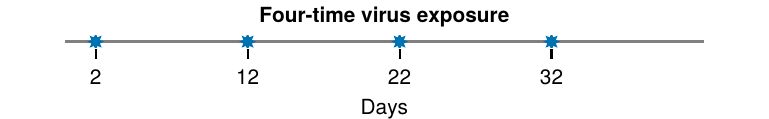}
        \setlength{\belowcaptionskip}{0cm}
        \caption*{}
        \label{fig:fourtimes9}
    \end{subfigure}
    \begin{subfigure}[t]{0.3\textwidth}
        \centering
        \includegraphics[width = 2in]{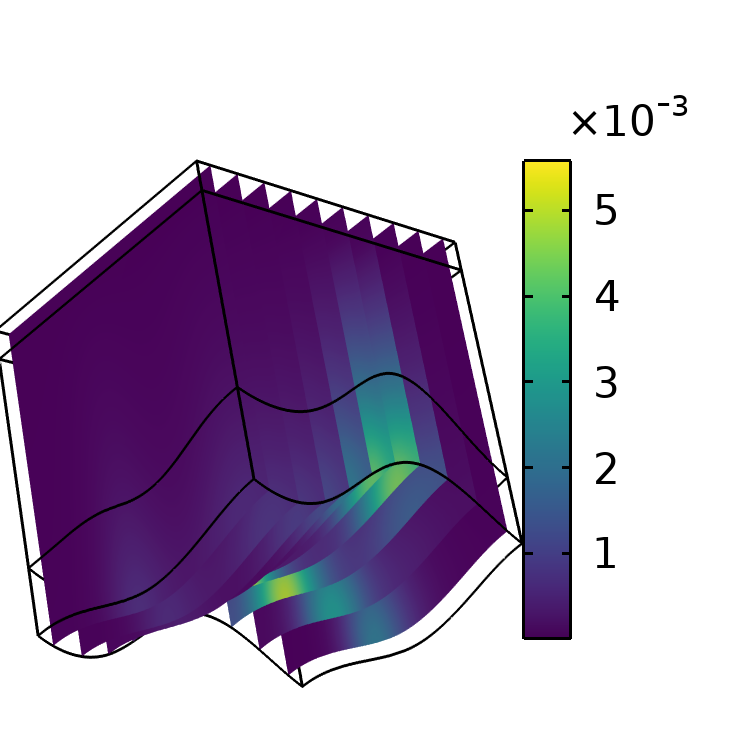}
        \caption{$t=100 \; \text{d}$}
        \label{fig:sim_times_4_100_a}
    \end{subfigure}
    \hspace*{\fill}%
    \begin{subfigure}[t]{0.3\textwidth}
        \centering
        \includegraphics[width = 2in]{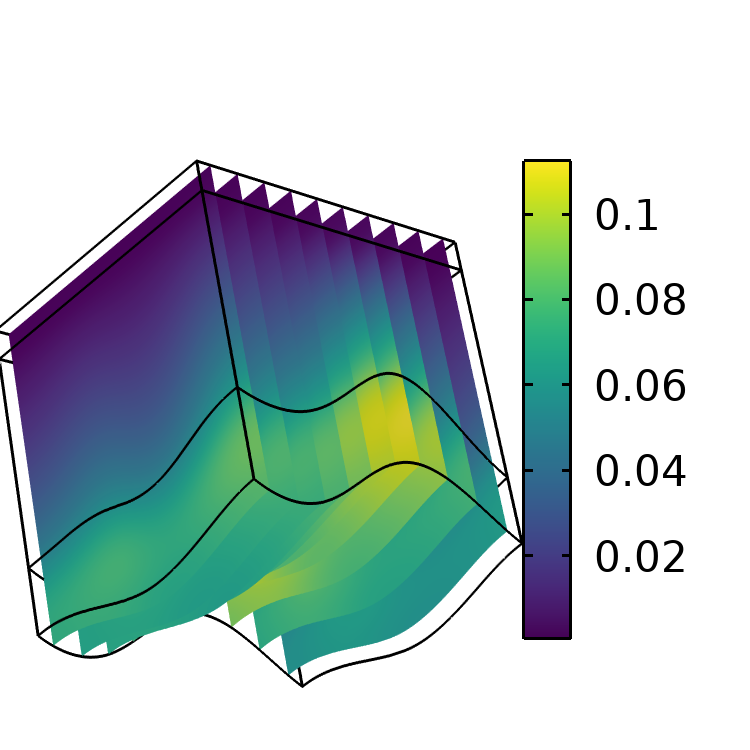}
        \caption{$t=800 \; \text{d}$}
        \label{fig:sim_times_4_800_b}
    \end{subfigure}
    \hspace*{\fill}%
    \begin{subfigure}[t]{0.3\textwidth}
        \centering
        \includegraphics[width = 2in]{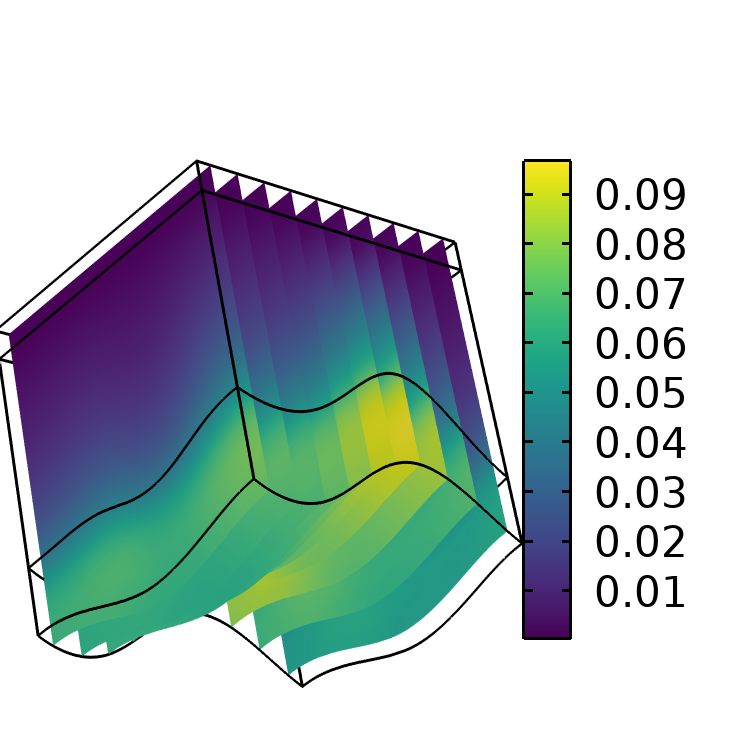}
        \caption{$t=1350 \; \text{d}$}
        \label{fig:sim_times_4_1350_c}
    \end{subfigure}
    \caption{Snapshots of dysplastic cells for three different days. Here $T_s=\bigcup\limits_{i\in\left\{2,12,22,32\right\}}{\left[i,i+\frac{1}{48}\right]}\;\text{d}$. We observe that a four-time virus exposure speeds up the establishment of dysplastic cells when compared to a two-time exposure, for the same parameter set.}
    \label{fig:times4}
\end{figure}


\section{Conclusion and discussion} \label{conclusion}
In this paper, we introduced a model of nonlinear partial differential equations describing the dynamics of epithelial cells, viral particles, dysplastic cells and immune cells. The model was then used to numerically investigate the key characteristics of dysplasia progression. From this investigation we deduced that 
\begin{itemize}
    \item the shape of the basement membrane alone can dictate the ability of dysplastic cells to establish themselves in the epithelium
    \item the smaller the flux of immune cells through the basement membrane, the greater the concentration of dysplastic cells
    \item the frequency of the viral exposure determines both the existence of dysplastic cells and the speed of their aggregation. 
\end{itemize}

Our work can be extended from both an analytical and a modelling point of view. Concerning the analytical perspective, we plan to study the model in terms of global well-posedness and stability analysis. Moreover, we intend to explicitly determine the dependence of the solution on the geometrical data of the membrane, the flux of immune cells, as well as the source of the viral exposure.

As for the modelling perspective, we could extend this study by extracting a condition for the prediction of the compliant/critical points of the basement membrane that would cause the transition from the CIS to cancer. 
In addition, we plan to extensively study the role of different immune cells, instead of grouping them together into one variable. Moreover, a straight-forward generalization includes the consideration of infection with different strains of HPV. Finally, we plan to incorporate a new dependent variable that will model the level of mutation of epithelial cells.
\section*{Acknowledgement}
The authors would like to thank Stratis Manifavas for his comments on random variables used in \hyperref[l1]{\S \ref*{l1}}.
\bibliographystyle{plainurl}
\bibliography{mybibfile}\label{bibliography}

\begin{thebibliography}{10}

\bibitem{asih2016dynamics}
Tri Sri~Noor Asih, Suzanne Lenhart, Steven Wise, Lina Aryati, F.~Adi-Kusumo, Mardiah~S. Hardianti, and Jonathan Forde.
\newblock {The dynamics of HPV infection and cervical cancer cells}.
\newblock {\em Bulletin of Mathematical Biology}, 78:4--20, 2016.
\newblock \href {https://doi.org/10.1007/s11538-015-0124-2} {\path{doi:10.1007/s11538-015-0124-2}}.

\bibitem{barnabas2006epidemiology}
Ruanne~V. Barnabas, P{\"a}ivi Laukkanen, Pentti Koskela, Osmo Kontula, Matti Lehtinen, and Geoff~P. Garnett.
\newblock {Epidemiology of HPV 16 and cervical cancer in Finland and the potential impact of vaccination: mathematical modelling analyses}.
\newblock {\em PLoS Medicine}, 3(5):e138, 2006.
\newblock \href {https://doi.org/10.1371/journal.pmed.0030138} {\path{doi:10.1371/journal.pmed.0030138}}.

\bibitem{barnsley1988science}
Michael~F. Barnsley, Robert~L. Devaney, Benoit~B. Mandelbrot, Heinz-Otto Peitgen, Dietmar Saupe, Richard~F. Voss, Yuval Fisher, and Michael McGuire.
\newblock {\em The Science of Fractal Images}, volume~1.
\newblock Springer, 1988.

\bibitem{blair2014can}
Karen~L Blair and Caroline~F Pukall.
\newblock Can less be more? comparing duration vs. frequency of sexual encounters in same-sex and mixed-sex relationships.
\newblock {\em The Canadian Journal of Human Sexuality}, 23(2):123--136, 2014.
\newblock \href {https://doi.org/10.3138/cjhs.2393} {\path{doi:10.3138/cjhs.2393}}.

\bibitem{brown2010hpv}
Victoria Brown and Katrin A.~J. White.
\newblock {The HPV vaccination strategy: could male vaccination have a significant impact?}
\newblock {\em Computational and Mathematical Methods in Medicine}, 11(3):223--237, 2010.
\newblock \href {https://doi.org/10.1080/17486700903486613} {\path{doi:10.1080/17486700903486613}}.

\bibitem{brown2011role}
Victoria~L. Brown and K.~A.~Jane White.
\newblock {The role of optimal control in assessing the most cost-effective implementation of a vaccination programme: HPV as a case study}.
\newblock {\em Mathematical Biosciences}, 231(2):126--134, 2011.
\newblock \href {https://doi.org/10.1016/j.mbs.2011.02.009} {\path{doi:10.1016/j.mbs.2011.02.009}}.

\bibitem{bzhalava2013systematic}
Davit Bzhalava, Peng Guan, Silvia Franceschi, Joakim Dillner, and Gary Clifford.
\newblock A systematic review of the prevalence of mucosal and cutaneous human papillomavirus types.
\newblock {\em Virology}, 445(1-2):224--231, 2013.
\newblock \href {https://doi.org/10.1016/j.virol.2013.07.015} {\path{doi:10.1016/j.virol.2013.07.015}}.

\bibitem{intgen2017}
Albert Einstein College of Medicine Analytical Biological Services et~al. Cancer Genome Atlas Research~Network.
\newblock Integrated genomic and molecular characterization of cervical cancer.
\newblock {\em Nature}, 543:378--384, 2017.
\newblock \href {https://doi.org/10.1038/nature21386} {\path{doi:10.1038/nature21386}}.

\bibitem{chakraborty2019role}
S.~Chakraborty, Xianbing Cao, S.~Bhattyacharya, and P.~K. Roy.
\newblock {The Role of HPV on cervical cancer with several functional response: a control based comparative study}.
\newblock {\em Computational Mathematics and Modeling}, 30:439--453, 2019.
\newblock \href {https://doi.org/10.1007/s10598-019-09469-4} {\path{doi:10.1007/s10598-019-09469-4}}.

\bibitem{chandrasoma2017gerd}
Parakrama~T. Chandrasoma.
\newblock {\em GERD: A New Understanding of Pathology, Pathophysiology, and Treatment}.
\newblock Academic Press, 2017.
\newblock \href {https://doi.org/10.1016/C2015-0-06626-6} {\path{doi:10.1016/C2015-0-06626-6}}.

\bibitem{chang2019beyond}
Julie Chang and Ovijit Chaudhuri.
\newblock Beyond proteases: Basement membrane mechanics and cancer invasion.
\newblock {\em Journal of Cell Biology}, 218(8):2456--2469, 2019.
\newblock \href {https://doi.org/10.1083/jcb.201903066} {\path{doi:10.1083/jcb.201903066}}.

\bibitem{demay1999practical}
Richard~M. DeMay.
\newblock {\em Practical Principles of Cytopathology}.
\newblock American Society of Clinical Pathologists Press, 1999.

\bibitem{doorbar2015human}
John Doorbar, Nagayasu Egawa, Heather Griffin, Christian Kranjec, and Isao Murakami.
\newblock Human papillomavirus molecular biology and disease association.
\newblock {\em Reviews in Medical Virology}, 25:2--23, 2015.
\newblock \href {https://doi.org/10.1002/rmv.1822} {\path{doi:10.1002/rmv.1822}}.

\bibitem{elbasha2008global}
Elamin~H. Elbasha.
\newblock {Global stability of equilibria in a two-sex HPV vaccination model}.
\newblock {\em Bulletin of Mathematical Biology}, 70:894--909, 2008.
\newblock \href {https://doi.org/10.1007/s11538-007-9283-0} {\path{doi:10.1007/s11538-007-9283-0}}.

\bibitem{guan2012human}
Peng Guan, Rebecca Howell-Jones, Ni~Li, Laia Bruni, Silvia De~Sanjos{\'e}, Silvia Franceschi, and Gary~M. Clifford.
\newblock {Human papillomavirus types in 115,789 HPV-positive women: a meta-analysis from cervical infection to cancer}.
\newblock {\em International Journal of Cancer}, 131(10):2349--2359, 2012.
\newblock \href {https://doi.org/10.1002/ijc.27485} {\path{doi:10.1002/ijc.27485}}.

\bibitem{gurmu2020mathematical}
E.~D. Gurmu, B.~K. Bole, and P.~R. Koya.
\newblock {Mathematical model for co-infection of HPV with cervical cancer and HIV with AIDS diseases}.
\newblock {\em International Journal of Scientific Research in Mathematical and Statistical Sciences}, 7(2):107--121, 2020.

\bibitem{hanahan2000hallmarks}
Douglas Hanahan and Robert~A. Weinberg.
\newblock The hallmarks of cancer.
\newblock {\em Cell}, 100(1):57--70, 2000.
\newblock \href {https://doi.org/10.1016/j.cell.2011.02.013} {\path{doi:10.1016/j.cell.2011.02.013}}.

\bibitem{keiffer2021recent}
Timothy~R. Keiffer, Sarah Soorya, and Martin~J. Sapp.
\newblock Recent advances in our understanding of the infectious entry pathway of human papillomavirus type 16.
\newblock {\em Microorganisms}, 9(10):2076, 2021.
\newblock \href {https://doi.org/10.3390/microorganisms9102076} {\path{doi:10.3390/microorganisms9102076}}.

\bibitem{kierszenbaum2015histology}
Abraham~L. Kierszenbaum and Laura Tres.
\newblock {\em Histology and Cell Biology: An Introduction to Pathology E-Book}.
\newblock Elsevier Health Sciences, 2015.

\bibitem{kurman2013blaustein}
Robert~J. Kurman.
\newblock {\em Blaustein's Pathology of the Female Genital tract}.
\newblock Springer Science \& Business Media, 2013.

\bibitem{kusakabe2023carcinogenesis}
Misako Kusakabe, Ayumi Taguchi, Kenbun Sone, Mayuyo Mori, and Yutaka Osuga.
\newblock Carcinogenesis and management of human papillomavirus-associated cervical cancer.
\newblock {\em International Journal of Clinical Oncology}, 28:965--974, 2023.
\newblock \href {https://doi.org/10.1007/s10147-023-02337-7} {\path{doi:10.1007/s10147-023-02337-7}}.

\bibitem{lee2007differentiation}
Choogho Lee and Laimonis~A. Laimins.
\newblock The differentiation-dependent life cycle of human papillomaviruses in keratinocytes.
\newblock In {\em The papillomaviruses}, pages 45--67. Springer, 2007.
\newblock \href {https://doi.org/10.1007/978-0-387-36523-7_4} {\path{doi:10.1007/978-0-387-36523-7_4}}.

\bibitem{lee2012mathematical}
Shernita~L. Lee and Ana~M. Tameru.
\newblock {A mathematical model of human papillomavirus (HPV) in the United States and its impact on cervical cancer}.
\newblock {\em Journal of Cancer}, 3:262, 2012.
\newblock \href {https://doi.org/10.7150/jca.4161} {\path{doi:10.7150/jca.4161}}.

\bibitem{ljubojevic2014hpv}
Suzana Ljubojevic and Mihael Skerlev.
\newblock {HPV-associated diseases}.
\newblock {\em Clinics in Dermatology}, 32(2):227--234, 2014.
\newblock \href {https://doi.org/10.1016/j.clindermatol.2013.08.007} {\path{doi:10.1016/j.clindermatol.2013.08.007}}.

\bibitem{marusyk2012intra}
Andriy Marusyk, Vanessa Almendro, and Kornelia Polyak.
\newblock Intra-tumour heterogeneity: a looking glass for cancer?
\newblock {\em Nature Reviews Cancer}, 12(5):323--334, 2012.
\newblock \href {https://doi.org/10.1038/nrc3261} {\path{doi:10.1038/nrc3261}}.

\bibitem{mendez2022revisiting}
Luis~Fernando M{\'e}ndez-L{\'o}pez.
\newblock Revisiting epithelial carcinogenesis.
\newblock {\em International Journal of Molecular Sciences}, 23(13):7437, 2022.
\newblock \href {https://doi.org/10.3390/ijms23137437} {\path{doi:10.3390/ijms23137437}}.

\bibitem{mescher2018junqueira}
Anthony~L. Mescher.
\newblock {\em Junqueira’s Basic Histology: Text and Atlas}.
\newblock New York: McGraw Hill, 2018.

\bibitem{moody2010human}
Cary~A. Moody and Laimonis~A. Laimins.
\newblock Human papillomavirus oncoproteins: pathways to transformation.
\newblock {\em Nature Reviews Cancer}, 10(8):550--560, 2010.
\newblock \href {https://doi.org/10.1038/nrc2886} {\path{doi:10.1038/nrc2886}}.

\bibitem{munger2004mechanisms}
Karl M\"unger, Amy Baldwin, Kirsten~M Edwards, Hiroyuki Hayakawa, Christine~L Nguyen, Michael Owens, Miranda Grace, and KyungWon Huh.
\newblock Mechanisms of human papillomavirus-induced oncogenesis.
\newblock {\em Journal of Virology}, 78(21):11451--11460, 2004.
\newblock \href {https://doi.org/10.1128/JVI.78.21.11451-11460.2004} {\path{doi:10.1128/JVI.78.21.11451-11460.2004}}.

\bibitem{nci2023hpv}
{National Cancer Institute}.
\newblock {HPV and Cancer}, 1 March 2019. Retrieved 27 April 2023.
\newblock URL: \url{https://www.cancer.gov/about-cancer/causes-prevention/risk/infectious-agents/hpv-and-cancer}.

\bibitem{ogino2011cancer}
Shuji Ogino, J{\'e}r{\^o}me Galon, Charles~S. Fuchs, and Glenn Dranoff.
\newblock Cancer immunology—analysis of host and tumor factors for personalized medicine.
\newblock {\em Nature Reviews Clinical Oncology}, 8(12):711--719, 2011.
\newblock \href {https://doi.org/10.1038/nrclinonc.2011.122} {\path{doi:10.1038/nrclinonc.2011.122}}.

\bibitem{park2023apoptosis}
Woo-Yong Park, Justin~M. Gray, Ronald~J. Holewinski, Thorkell Andresson, Jae~Young So, Carmelo Carmona-Rivera, M.~Christine Hollander, Howard~H. Yang, Maxwell Lee, Mariana~J. Kaplan, et~al.
\newblock Apoptosis-induced nuclear expulsion in tumor cells drives {S100a4}-mediated metastatic outgrowth through the {RAGE} pathway.
\newblock {\em Nature Cancer}, 4(3):419--435, 2023.
\newblock \href {https://doi.org/10.1038/s43018-023-00524-z} {\path{doi:10.1038/s43018-023-00524-z}}.

\bibitem{pastar2014epithelialization}
Irena Pastar, Olivera Stojadinovic, Natalie~C. Yin, Horacio Ramirez, Aron~G. Nusbaum, Andrew Sawaya, Shailee~B. Patel, Laiqua Khalid, Rivkah~R. Isseroff, and Marjana Tomic-Canic.
\newblock Epithelialization in wound healing: a comprehensive review.
\newblock {\em Advances in Wound Care}, 3(7):445--464, 2014.
\newblock \href {https://doi.org/https://10.1089/wound.2013.0473} {\path{doi:https://10.1089/wound.2013.0473}}.

\bibitem{perkins2023cervical}
Rebecca~B. Perkins, Nicolas Wentzensen, Richard~S. Guido, and Mark Schiffman.
\newblock Cervical cancer screening: a review.
\newblock {\em Jama}, 330(6):547--558, 2023.
\newblock \href {https://doi.org/10.1001/jama.2023.13174} {\path{doi:10.1001/jama.2023.13174}}.

\bibitem{pevsut2021human}
Ena Pe{\v{s}}ut, Anamaria {\DJ}uki{\'c}, Lucija Luli{\'c}, Josipa Skelin, Ivana {\v{S}}imi{\'c}, Nina Milutin~Ga{\v{s}}perov, Vjekoslav Tomai{\'c}, Ivan Sabol, and Magdalena Grce.
\newblock Human papillomaviruses-associated cancers: an update of current knowledge.
\newblock {\em Viruses}, 13(11):2234, 2021.
\newblock \href {https://doi.org/10.3390/v13112234} {\path{doi:10.3390/v13112234}}.

\bibitem{pozzi2017nature}
Ambra Pozzi, Peter~D Yurchenco, and Renato~V. Iozzo.
\newblock The nature and biology of basement membranes.
\newblock {\em Matrix Biology}, 57:1--11, 2017.
\newblock \href {https://doi.org/https://10.1016/j.matbio.2016.12.009} {\path{doi:https://10.1016/j.matbio.2016.12.009}}.

\bibitem{prendiville2017colposcopy}
Walter Prendiville and Rengaswamy Sankaranarayanan.
\newblock {\em Colposcopy and Treatment of Cervical Precancer}.
\newblock International Agency for Research on Cancer, World Health Organization, 2017.

\bibitem{rajan2023mathematical}
Praveen~Kumar Rajan, Murugesan Kuppusamy, and Oluwaseun~F. Egbelowo.
\newblock {A mathematical model for human papillomavirus and its impact on cervical cancer in India}.
\newblock {\em Journal of Applied Mathematics and Computing}, 69(1):753--770, 2023.
\newblock \href {https://doi.org/10.1007/s12190-022-01767-2} {\path{doi:10.1007/s12190-022-01767-2}}.

\bibitem{sado2019mathematical}
Abdulsamad~Engida Sado.
\newblock {Mathematical modeling of cervical cancer with HPV transmission and vaccination}.
\newblock {\em Science Journal of Applied Mathematics and Statistics}, 7(2):21--25, 2019.
\newblock \href {https://doi.org/10.11648/j.sjams.20190702.13} {\path{doi:10.11648/j.sjams.20190702.13}}.

\bibitem{salcedo2021intraepithelial}
Mila~Pontremoli Salcedo, Natacha Phoolcharoen, and Kathleen~M. Schmeler.
\newblock Intraepithelial neoplasia of the lower genital tract (cervix, vagina, vulva): etiology, screening, diagnosis, management.
\newblock In {\em Comprehensive Gynecology}, page 637. Elsevier, 2021.
\newblock \href {https://doi.org/10.1016/B978-0-323-65399-2.00038-3} {\path{doi:10.1016/B978-0-323-65399-2.00038-3}}.

\bibitem{santesso2016world}
Nancy Santesso, Reem~A Mustafa, Holger~J Sch{\"u}nemann, Marc Arbyn, Paul~D Blumenthal, Joanna Cain, Michael Chirenje, Lynette Denny, Hugo De~Vuyst, Linda~O'Neal Eckert, et~al.
\newblock World health organization guidelines for treatment of cervical intraepithelial neoplasia 2--3 and screen-and-treat strategies to prevent cervical cancer.
\newblock {\em International Journal of Gynecology \& Obstetrics}, 132(3):252--258, 2016.
\newblock \href {https://doi.org/10.1016/j.ijgo.2015.07.038} {\path{doi:10.1016/j.ijgo.2015.07.038}}.

\bibitem{schiffman2007human}
Mark Schiffman, Philip~E. Castle, Jose Jeronimo, Ana~C. Rodriguez, and Sholom Wacholder.
\newblock Human papillomavirus and cervical cancer.
\newblock {\em The Lancet}, 370(9590):890--907, 2007.
\newblock \href {https://doi.org/10.1016/S0140-6736(07)61416-0} {\path{doi:10.1016/S0140-6736(07)61416-0}}.

\bibitem{schiller2010current}
John~T. Schiller, Patricia~M. Day, and Rhonda~C. Kines.
\newblock Current understanding of the mechanism of {HPV} infection.
\newblock {\em Gynecologic Oncology}, 118(1):S12--S17, 2010.
\newblock \href {https://doi.org/10.1016/j.ygyno.2010.04.004} {\path{doi:10.1016/j.ygyno.2010.04.004}}.

\bibitem{sjodin2017generate}
Bjorn Sjodin.
\newblock How to generate random surfaces in {COMSOL Multiphysics}{\textregistered}.
\newblock {\em Comsol Blog}, 2017.

\bibitem{solis2017numerical}
Francisco~J. Solis and Luz~M. Gonzalez.
\newblock A numerical approach for a model of the precancer lesions caused by the human papillomavirus.
\newblock {\em Journal of Difference Equations and Applications}, 23(6):1093--1104, 2017.
\newblock \href {https://doi.org/10.1080/10236198.2017.1318858} {\path{doi:10.1080/10236198.2017.1318858}}.

\bibitem{solis2023nonlinear}
Francisco~J. Solis and Luz~M. Gonzalez.
\newblock {A nonlinear transport--diffusion model for the interactions between immune system cells and HPV-infected cells}.
\newblock {\em Nonlinear Dynamics}, pages 1--15, 2023.
\newblock \href {https://doi.org/10.1007/s11071-023-08616-2} {\path{doi:10.1007/s11071-023-08616-2}}.

\bibitem{stubenrauch1999human}
Frank Stubenrauch and Laimonis~A. Laimins.
\newblock Human papillomavirus life cycle: active and latent phases.
\newblock In {\em Seminars in Cancer Biology}, volume~9, pages 379--386. Elsevier, 1999.
\newblock \href {https://doi.org/10.1006/scbi.1999.0141} {\path{doi:10.1006/scbi.1999.0141}}.

\bibitem{sung2021global}
Hyuna Sung, Jacques Ferlay, Rebecca~L. Siegel, Mathieu Laversanne, Isabelle Soerjomataram, Ahmedin Jemal, and Freddie Bray.
\newblock Global cancer statistics 2020: {GLOBOCAN} estimates of incidence and mortality worldwide for 36 cancers in 185 countries.
\newblock {\em CA: A Cancer Journal for Clinicians}, 71(3):209--249, 2021.
\newblock \href {https://doi.org/10.3322/caac.21660} {\path{doi:10.3322/caac.21660}}.

\bibitem{tommasino2014human}
Massimo Tommasino.
\newblock The human papillomavirus family and its role in carcinogenesis.
\newblock In {\em Seminars in Cancer Biology}, volume~26, pages 13--21. Elsevier, 2014.
\newblock \href {https://doi.org/10.1016/j.semcancer.2013.11.002} {\path{doi:10.1016/j.semcancer.2013.11.002}}.

\bibitem{veldhuijzen2010factors}
Nienke~J. Veldhuijzen, Peter J.~F. Snijders, Peter Reiss, Chris J. L.~M. Meijer, and Janneke H. H~.M. van~de Wijgert.
\newblock Factors affecting transmission of mucosal human papillomavirus.
\newblock {\em The Lancet Infectious Diseases}, 10(12):862--874, 2010.
\newblock \href {https://doi.org/10.1016/S1473-3099(10)70190-0} {\path{doi:10.1016/S1473-3099(10)70190-0}}.

\bibitem{walboomers1999human}
Jan M.~M. Walboomers, Marcel~V. Jacobs, M.~Michele Manos, F.~Xavier Bosch, J.~Alain Kummer, Keerti~V. Shah, Peter J.~F. Snijders, Julian Peto, Chris J. L.~M. Meijer, and Nubia Mu{\~n}oz.
\newblock Human papillomavirus is a necessary cause of invasive cervical cancer worldwide.
\newblock {\em The Journal of Pathology}, 189(1):12--19, 1999.
\newblock \href {https://doi.org/10.1002/(SICI)1096-9896(199909)189:1<12::AID-PATH431>3.0.CO;2-F} {\path{doi:10.1002/(SICI)1096-9896(199909)189:1<12::AID-PATH431>3.0.CO;2-F}}.

\bibitem{Whohc}
{World Health Organization}.
\newblock {Histopathology of the uterine cervix - digital atlas: WHO histological classification of tumours of the uterine cervix}.
\newblock URL: \url{https://screening.iarc.fr/atlasclassifwho.php}.

\bibitem{Who2030}
{World Health Organization}.
\newblock {Reaching 2030 cervical cancer elimination targets - New WHO recommendations for screening and treatment of cervical pre-cancer}, 2021.
\newblock URL: \url{https://www.who.int/news-room/events/detail/2021/07/06/default-calendar/reaching-2030-cervical-cancer-elimination-targets}.

\bibitem{Who2022vaccine}
{World Health Organization}.
\newblock {Human papillomavirus vaccines (HPV)}, 2022.
\newblock URL: \url{https://www.who.int/teams/immunization-vaccines-and-biologicals/diseases/human-papillomavirus-vaccines-(HPV)}.

\bibitem{Who2023cc}
{World Health Organization}.
\newblock Cervical cancer, 2023.
\newblock URL: \url{https://www.who.int/news-room/fact-sheets/detail/cervical-cancer}.

\bibitem{mondiale2022human}
{World Health Organization--Organisation mondiale de la Santé}.
\newblock {Human papillomavirus vaccines: WHO position paper (2022 update)--Vaccins contre les papillomavirus humains: note de synth{\`e}se de l' OMS (mise {\`a} jour de 2022)}.
\newblock {\em Weekly Epidemiological Record--Relev{\'e} {\'E}pid{\'e}miologique Hebdomadaire}, 97(50):645--672, 2022.


\bibitem{comsol}
{COMSOL AB}.
\newblock COMSOL Multiphysics $\circledR$.
\newblock URL: \url{https://comsol.com}.



\end{thebibliography}
\addcontentsline{toc}{chapter}{Bibliography}
\end{document}